 \documentclass[utf8]{FrontiersinHarvard}

\usepackage{url,hyperref,lineno,microtype,subcaption}
 \usepackage[onehalfspacing]{setspace}

 \usepackage{hyperref}  
 \usepackage{graphicx}
 \usepackage{amsmath}
 \usepackage{amssymb}
 \usepackage{yfonts}
 \usepackage{slashed} 
 \usepackage{marvosym}
 \usepackage{wasysym}
 \usepackage{xcolor}
 \usepackage{mathrsfs}
 \usepackage{amsbsy}

 \newcommand{\EAE }{{{EAE\hspace{1.0mm}}}}
 \newcommand{\EAEc}{{{EAE}}}
\newcommand{\EAEd}{{{EAE}}}
 
 \newcommand{\gx}{{\rm gx}}
\newcommand{\dS}{{\rm dS}}

\newcommand{\rmi}{{\rm i}}

\newcommand{\gn}{h}

\newcommand{\varrhoq}{j}

\newcommand{\inh}{{\rm i}}

\newcommand{\dele}{\delta_e}
\newcommand{\Ome}{\Omega_e}

\newcommand{\Sn}{N}
\newcommand{\Se}{S}
\newcommand{\bSe}{\bar S}

\newcommand{\bh}{{\rm bh}}

\renewcommand{\em}{{{\rm em}\,}}

\newcommand{\rmb}{{\rm b}}

\newcommand{\statV}{{\rm statV}}

\newcommand{\sig}{\sigma}
 
\newcommand{\vE}{{\bf E}}

 \newcommand{\veps}{\boldsymbol{\varepsilon}}

\newcommand{\vr}{{\bf r}}  
  
\newcommand{\vR}{{\bf R}}

\newcommand{\heen}{{(1)}}

\newcommand{\cmb}{{\rm cmb}}

\newcommand{\now}{{\rm now}}

\newcommand{\qad}{\hspace{1mm}}

\newcommand{\alp}{\alpha}

\newcommand{\Gam}{\Gamma}

\newcommand{\lam}{\lambda}
\newcommand{\Lam}{\Lambda}

\newcommand{\km}{{\rm km}}
\newcommand{\kpc}{{\rm kpc}}
\newcommand{\Mpc}{{\rm Mpc}}

\newcommand{\bcg}{{\rm bcg}}

\newcommand{\rmm}{{\rm m}}
\newcommand{\rms}{{\rm s}}

\newcommand{\co}{{\rm co}}

\newcommand{\rhoA}{\rho_E}

\newcommand{\ezp}{{\it e}}
\newcommand{\zpe}{\lam}

\newcommand{\dm}{{\rm dm}}

\newcommand{\rot}{{\rm rot}}

\newcommand{\mmin}{{\hspace{0.2mm}-\hspace{0.2mm}}}
\newcommand{\pplus}{{\hspace{0.2mm}+\hspace{0.2mm}}}

\newcommand{\se}{{\rm se}}

\newcommand{\refl}{\ref}

\renewcommand{\d}{{\rm d}}

\newcommand{\cm}{{\rm cm}}

\newcommand{\itf}{{\it f}}

\newcommand{\diag}{{\rm diag}}

\newcommand{\tot}{{\rm tot}}
\newcommand{\gr}{{\rm gr}}

\newcommand{\half}{{\frac{1}{2}}}

\renewcommand{\min}{{\rm min}}
\renewcommand{\max}{{\rm max}}

\newcommand{\myskip}[1]{}

\newcommand{\vk}{{\bf k}}

\newcommand{\vH}{{\bf H}}  
\newcommand{\vJ}{{\bf J}}  
\newcommand{\vL}{{\bf L}}  
\newcommand{\vP}{{\bf P}}

\newcommand{\vp}{{\bf p}} 
\newcommand{\onedot}{\,\,\,}  

\newcommand{\ed}{\onedot}
\newcommand{\ednu}{{\onedot\nu}}

\newcommand{\Om}{\Omega}
\newcommand{\BEQ}{\begin{eqnarray}}   
\newcommand{\EEQ}{\end{eqnarray}}   
\newcommand{\BEA}{\begin{eqnarray}}   
\newcommand{\EEA}{\end{eqnarray}}   
\newcommand{\nn}{\nonumber }   
\renewcommand{\d}{{\rm d}}   
\newcommand{\del}{\delta}

\newcommand{\vth}{{\rm m}}   
   
\newcommand{\eps}{\varepsilon}   
\newcommand{\om}{\omega}

\newcommand{\mn}{{\mu\nu}}

\renewcommand{\diag}{{\rm diag}}

\newcommand{\cal}{\mathcal}

\newcommand{\cC}{{ C}}

\newcommand{\cF}{{\cal F}}
\newcommand{\cFE}{{\cal F}_E}
\newcommand{\cFG}{{\cal F}_G}

\newcommand{\cO}{{\cal O}}

\def\keyFont{\fontsize{8}{11}\helveticabold }
\def\firstAuthorLast{T.M. Nieuwenhuizen} 
\def\Authors{Theodorus Maria Nieuwenhuizen}
\def\Address{Institute for Theoretical Physics,  University of Amsterdam 
 \\ Science Park 904, 1090 GL  Amsterdam, The Netherlands} 
\def\corrAuthor{T.M. Nieuwenhuizen}

\def\corrEmail{t.m.nieuwenhuizen@uva.nl}

\begin{document}
\onecolumn
\firstpage{1}

\title[Solution of the dark matter riddle  within standard model physics]{Solution of the dark matter riddle  within standard model physics:
From black holes, galaxies and clusters to cosmology} 

\author[\firstAuthorLast ]{\Authors} 
\address{} 
\correspondance{} 

 \extraAuth{}

 \maketitle

\begin{abstract}
\section{}
It is postulated that the energy density of the (quantum) vacuum acts firstly as dark energy and secondly as a part of dark matter. 
Assisted by electric fields arising from a small charge mismatch in the cosmic plasma, it can condense on mass concentrations.
No longer participating in the cosmic expansion, this constitutes  
``electro-aether-energy'' (\EAEc),
``electro-zero-point-energy''  or   ``electro-vacuum-energy'',  
which solves the dark matter riddle without new physics. 
 A radial electric field of 1 kV/m is predicted in the Galaxy.
For proper electric fields, \EAE   can cover the results deduced with MOND. An instability allows a speedy filling of dark matter cores.
Hydrostatic equilibrium in galaxy clusters is obeyed.  Flowing in of  
\EAE   explains why black holes become supermassive, do not have mass gaps  and overcome the final parsec problem.
Rupture of charged clouds reduces, e. g.,  the primordial baryon cloud to the cosmic web.
The large coherence scale of the electric field acts as a scaffold for gentle galaxy formation and  their vast polar structures. 
In galaxy merging and bars, there occurs no dynamical friction. 
At cosmological scales,  \EAE  acts as pressureless dark matter. 
Its amount increases in time, which likely solves the Hubble tension by its late time physics.  A big crunch can occur.
Of the large cosmological constant injected at the Big Bang, a small part kept that form, without fine-tuning.  


\tiny
 \keyFont{ \section{Keywords:} dark matter, dark energy, standard model, vacuum, zero point energy, aether, Hubble tension, 
 early structure formation}
\end{abstract}


\newpage

\section*{Table of contents}
\noindent\noindent
 1. Introduction \\
 1.1 Short history of dark matter \\
 1.2 Present status \\
 1.3 A new approach to the problem of dark matter \\
 1.4 Nomenclature for the proposed form of dark matter \\
 1.5 Setup

\noindent\noindent
 2 THE STANDARD MODEL AND ITS ZERO POINT ENERGY \\
 2.1 The essence: the Casimir effect \\
 2.2 Zero point/aether energy in the black hole interior \\
 2.3 Zero point energy as a physical entity \\
 2.4 Zero point energy as the dark energy \\
 2.5 Electric field as partner in zero point energy condensation 

\noindent\noindent
3 THEORETICAL FRAMEWORK \\ 
3.1 Full Einstein equations \\ 
3.2 Coulomb electrostatics in electrogravity \\
3.3 Linearized Einstein equations \\
3.4 Electro-zero-point energy (electro-aether energy) as the dark matter \\
3.5 A toy galaxy \\
3.6 The matter components  \\
3.7 The X-ray gas \\
3.8 Hydrostatic equilibrium

\noindent\noindent
4 PHYSICAL ESTIMATES \\
4.1 Charge mismatch in the plasma \\
4.2 Galactic electric field \\
4.3 Seeding of galactic magnetic fields \\ 
4.4 Implementation of hydrostatic stability \\
4.5 Metastability vs instability \\

\noindent\noindent
5 ELECTRO-AETHER-ENERGY IN BLACK HOLES \\ 
5.1 Black hole metrics with a macroscopic core \\
5.2 Super-Eddington accretion by aether energy inflow \\
5.3 The final parsec problem under EAE accretion \\

\noindent\noindent
6 ELECTRO-AETHER-ENERGY IN GALAXIES \\
6.1 Relation to MOND \\ 
6.2 Stability analysis \\
6.3 An instability leading to explosive core growth \\
6.4 Evidence for constant-density, non-cusped cores \\
6.5 Dissolution of galactic cores \\
6.6 The electric field scaffold and the vast polar structure  \\

\noindent\noindent
7 ELECTRO-AETHER-ENERGY IN CLUSTERS \\
7.1 Modified isothermal sphere as a fit for lensing \\
7.2 The hydrostatic equilibrium puzzle in clusters \\
7.2.1 Hydrostatic equilibrium in $\Lambda$CDM \\
7.2.2 Hydrostatic equilibrium in EAE

\noindent\noindent
8 ELECTRO-AETHER-ENERGY IN COSMOLOGY \\
8.1 Zero pressure EAE equation of state \\
8.2 The Hubble tension and the increasing amount of dark matter \\
8.3 Friedmannology for aether energy condensation \\
8.4 A fit to CMB data \\
8.5 The large Hubble constant and the age of the Universe \\
8.6 The lopsidedness of the cosmos and the axis of evil \\
8.7 Beyond present \\
8.7.1 Black holes and the big crunch \\
8.8 Times near the Big Bang \\ 
8.8.1 High zero-point energy initial state: Automatic inflation

\noindent\noindent
9 CONCLUSION

\noindent\noindent
10 SUMMARY

\noindent\noindent
11 OUTLOOK

\noindent\noindent
12 ACKNOWLEDGEMENTS

\noindent\noindent
REFERENCES

\newpage

\section{Introduction}

\hfill {\it Dedicated to the memory of my Ph.D. advisor}, 

\hfill {\it Theodorus Wilhelmus Ruijgrok, 1927-2022.}

\vspace{3mm}

\subsection{Short history of dark matter}

The matter in the world and skies we experience is called  ``normal matter'' by specialists.
It consists of protons and neutrons, that are bound by the strong force in atomic nuclei, and electrons
that encircle the nucleus due to the Coulomb attraction between the positive protons and the negative electrons.

Nowadays it is understood, however, that normal matter makes up only some 5\% of the total mass budget 
in the Universe. In fact, the stars that we observe in the night sky make up only a modest 4 $\permil$ of the total  \citep{rich2009fundamentals};
most of the 5\% lies in hydrogen clouds and hydrogen bridges between galaxies, which can be observed 
in the 21 cm radio line due to spin flipping in the hydrogen nucleus, as was discovered by Hendrik van de Hulst  \citep{cook2001hendrik}.

The remaining 95\% of the total is matter that we do not perceive directly, even though its existence has been established rather firmly.
After long suspicion of dark stars and the suggestion of dark matter based on stellar velocities by  Jacobus Kapteyn in 1922 \citep{kapteyn2013First}, 
 the existence of dark matter was established by Fritz Zwicky in 1933  \citep{zwicky1933rotverschiebung}
and  much support for it has emerged. An account of the history of dark matter is given in \citep{de2017dark}. 
Actually,  this matter is not dark but transparent.
Its French name ``mati\`ere obscure'' translates as ``obscure matter'', hidden or unexplained matter, which does more justice to its nature.

Observations by Vera Rubin and Kent Ford in the seventies demonstrated that in the outer part of galaxies, circular orbits have 
nearly the same rotation speed \citep{rubin1970rotation}, constituting ``flat rotation curves". 
This has led to general acceptance of dark matter's existence.

In 1998, it was established that there also exists dark energy  \citep{riess1998observational,perlmutter1999measurements},
which constitutes some 70\% of the mass budget in the Universe, while dark matter makes up some 25\%.
Of all the matter/energy in the Universe, 95\% is unexplained so far; it will be the focus of this work.

\subsection{Present status}

Dark matter must be cold, that is:  slowly moving, in order to account for the creation of galaxies.
Many searches have been carried out.
The historical candidate, MACHOs, massive astrophysical compact halo objects such as dark stars or planets, 
has been ruled out as the main contributor  \citep{tisserand2007limits,alcock2000macho}.
 The next candidate is the WIMP (weakly interacting massive particle)
 proposed by Jim Peebles and followers \citep{peebles1982large,bond1982formation,blumenthal1982galaxy,blumenthal1984formation}.
 Being massive and moving slowly, it is termed ``cold'' and leads to the present paradigm of Lambda cold dark matter ($\Lam$CDM). 
 Intensive searching for possible candidates has not yielded a discovery \citep{arcadi2018waning}, 
 and its 10 years window since 2010, during which various new searches should find it if it exists, has all but closed \citep{bertone2010particle}.
Nowadays, the focus is shifting to axions and axion-like-particles \citep{wilczek1978problem,weinberg1978new,sikivie1983experimental},
 to warm DM \citep{boyarsky2014unidentified}, and to dark photons, see, e.g., \citep{pignol2015constraints}.

Of the many other approaches we mention theories without a new particle, like Modified Newtonian Dynamics (MOND)  \citep{milgrom1983modification}
and entropic \citep{verlinde2011origin} and emergent gravity \citep{verlinde2017emergent}, 
which leads to related predictions \citep{nieuwenhuizen2017zwicky,brouwer2017first}. Here the Newton law is modified for weak acceleration.
A definite test of general relativity versus extended theories of gravity has been worked out for gravitational wave observation \citep{corda2009interferometric}.

The present paradigm, $\Lambda$CDM, is quite successful, among others for its widely employed Navarro-Frenk-White (NFW) profile \citep{navarro1997universal}. 
Numerical codes for it are well developed, e.g. \citep{vogelsberger2014introducing}.
Workers in the field have time and again achieved to model apparent ``non-$\Lam$CDM'' aspects within the theory.
However, $\Lambda$CDM remains loaded with issues;
for reviews of challenges, see e.g.  \citep{kroupa2012dark,bull2016beyond,bullock2017small,perivolaropoulos2022challenges}.

The dark matter (DM) problem is still an outstanding riddle, but various aspects challenge $\Lam$CDM:
absence of dynamical friction in galaxy merging \citep{kroupa2015galaxies,oehm2024relevance} and in galactic bars \citep{roshan2021fast}; 
as per MOND, structures in the baryons correspond to structures in the rotation velocity.
The Hubble tension expresses the difference between the Hubble constant $H_0\approx 68$ km/s Gpc 
from the cosmic microwave radiation \citep{aghanim2020planck}  
versus the  73 km/s Gpc from supernovae in the nearby Universe \citep{brout2022pantheon}.
This is not solvable (within $\Lam$CDM) with new physics at low redshift \citep{keeley2023ruling}.
Likewise, the Lithium-7 problem expressing a factor $\sim 3$ larger nucleosynthesis 
prediction than its observed Spite plateau \citep{spite1982abundance,coc2005lithium,pitrou2018precision},
has remained open. 

Observation by the James Webb telescope poses further challenges for $\Lambda$CDM.
Black holes became supermassive early on. 
The   black hole ULAS J1342+0928 has a mass of $8\,10^8M_\odot$ at redshift $z=7.54$ \citep{banados2018800}. 
The one of GN-z11, an exceptionally luminous galaxy at $z=10.6$,
weighs several million solar masses and has accreted at about 5 times the Eddington rate \citep{maiolino2024small}. 
Early galaxies do not look chaotic but are  more mature and heavier than predicted.
A set of 12  of quiescent galaxies at redshifts $3-4$ has masses in the $10^{11}M_\odot$ range,
comparable to massive galaxies in the local Universe, 
and are quenched for more than a billion years \citep{nanayakkara2024population}.
For redshifts $z=$ 0.5 -- 8, many dwarf galaxies  are prolate \citep{van2014geometry,zhang2019evolution}, 
now termed ``bananas'' \citep{pandya2024galaxies}.
The current record holder is JADES-GS-z13-0, a galaxy with spectroscopically confirmed redshift of 13.2, 
which we observe as it existed  350 million years after the Big Bang \citep{curtis2023spectroscopic}.
As to clusters, an overdensity related to a  galaxy protocluster is observed already at redshift $z=7.88$ \citep{hashimoto2023reionization}.

These challenges put forward to abandon $\Lam$CDM and start anew.

\subsection{A new approach to the problem of dark matter}

We propose an approach\footnote{This manuscript is an update of arXiv:2303.04637, submitted to the arXiv on February 23, 2023.}
 towards many  such riddles {\it without new physics}; it will suffice to take
a new look at the {\it  vacuum} of the quantum field theory for the Standard Model of particle physics.
Let us consider an analogy with the atmosphere. A ``particle'' in the atmosphere can be identified with a tornado. 
Without ``particles'', the atmosphere is in its ``vacuum'' state.
But this is not a trivial state:  the weather can be gentle or involve winds, rains, storms, \ldots We shall take this notion over to the (quantum) vacuum, 
a state without matter, but not without energy. 
Because of Einstein's relation $E=mc^2$, vacuum energy condensed on galaxies has similar (but not the same) 
gravitational effects as  the purported particle dark matter.

Rather than invoking a new particle, we postulate that the zero point energy density
of the quantum aether has specific properties: 
next to being uniform and acting as  dark energy, it can be inhomogeneous and flow; 
it can condense on mass concentrations when assisted by electric fields; locally, it can be positive or negative.

Alternatively, we may view this vacuum as a classical aether (though not the historic fixed aether) with the said properties of energy.
Either way, the cosmological ``constant'' can slowly vary in space, and in time, while its local value can have either sign.
The gradient of the related negative or positive pressure acts as a genuine force density that can counteract, e.g.,  the electrostatic force density.
In this picture, {\it  dark energy partly acts},  in combination {\it with electrostatics}, as {\it the dark matter.}

\subsection{Nomenclature for the proposed form of dark matter}

For our theory of dark matter as a combination of { electrostatic energy} and {zero-point or aether energy}, 
 (or ``ether of space'', an old term from Sir Oliver Lodge), we have initially employed the term ``{\it electro-zero-point energy}'' (EZPE)
 \citep{nieuwenhuizen2023solution}. 
The drawback is that this refers to a quantum concept, while such a connection is, for now, merely an assumption.
A more neutral term is ``{\it electro-vacuum energy}''  (EVE) or ``{\it electro-aether energy}'' {EAE}.
Another one is  (the energy density related to) a {\it local cosmological constant} (LCC), a term we employed in 
our earlier work on black holes with a smooth (singularity-free) interior \citep{nieuwenhuizen2023exact}.
 
We are thus hampered by  traditions for the use the term ``vacuum'' as energetically either empty or non-empty.
Wishing to distinguish the standard vacuum of quantum field theory, we propose to call that ``the vacuum'', 
and use ``aether'' for the (substance that sustains a) nontrivial vacuum, one with nonzero energy density and/or flow.
Henceforward, we will use the acronym \EAEd.

\subsection{Setup}

The setup of this paper is as follows.
Aspects of zero point energy are discussed in section \ref{SM-ZPE}.
The theoretical framework is presented in section \ref{Fieldeqs}. 
The sizes of various effects are estimated in section \ref{phys-estimates}.
Applications to black holes are discussed in  section \ref{secBHs}.
The working in galaxies and comparison with MOND is treated in section \ref{EVE-gals}.
Section \ref{EVE-clusters} analyses the application to galaxy clusters.
An implementation for cosmology in various epochs is worked out in section \ref{EVE-cosmology}.
The work closes with a conclusion, a summary  and an outlook in sections \ref{conclusion}, \ref{summary} and \ref{outlook}.

\section{The standard model and its zero point energy}
\label{SM-ZPE}

The standard model of particle physics (SMPP) is a quantum field theory for the U(1) $\times$ SU(2)$\times$ SU(3) gauge group 
with three families of quarks and leptons. Since its conception in the 1960's and 1970's, all its particles have been established,
the latest ones being the top quark in 1995 and the Higgs boson in 2012.  Like in the decades before, 
SMPP has been capable to explain all experiments so far, 
the latest success being to rule out the breaking of lepton universality which was suggested in earlier experiments \citep{lhcb2022measurement}.

In quantum mechanics there is the notion of zero point energy (ZPE). A harmonic oscillator has a ZPE of $\half\hbar \omega$, where 
its oscillation frequency reads $\omega=\sqrt{k/m}$ with $k$ the spring constant and $m$ the mass.  
With $n$ quanta, it has energy levels $(n+\half)\hbar\om$.  While the energy differences  between the levels have
clear experimental meaning, the meaning of ZPE is less obvious.
Adiabatically changing either $k$ or $m$, changes the ZPE; the difference has a physical meaning.

A quantum field decomposes in a large set of  harmonic oscillators. Their total ZPE is formally a sum of $\pm\half\hbar\om_\vk$
 ($+$ for bosons, $-$ for fermions) over the $3d$ momenta $\vk$, a quartically divergent expression. If the momenta are
cut off at the Planck scale, there remains a result which is about 123 orders of magnitude\footnote{Other sources report 120 orders of magnitude;
both are estimates. The factor $10^{123}$ is sometimes called the Penrose number, named after the Nobel laureate Roger Penrose.}
larger than the dark energy  density in the present Universe. 
This enormous mismatch is a ground for unease to connection with ZPE. 

The question we put forward is however: Isn't there a more prominent role for the ZPE? 
Haven't we, by focussing on the particles and taking the ZPE for granted, been picking out the raisins while overlooking the pudding?
Is there room within the standard model to address its deficiencies like the description of the baryon asymmetry, dark matter and the 
dark energy related to the Universe's accelerated expansion?

\subsection{The essence: the Casimir effect}

In 1948 Hendrik Casimir discovered that two parallel conducting plates of area $A$ at distance $d$ have an energy $-\pi^2\hbar c A/720 d^3$
 \citep{casimir1948attraction,milton2003casimir,balian2004geometry}.
It is generally understood that this energy is gained from the quantum aether when bringing the plates from infinity to distance $d$.
Interpretations of the effect differ, however, since energy can not be localized in electrodynamics and neither in gravitation. 

Upon adiabatically moving the plates to a distance $d'<d$ (or $d'>d$), the aether energy changes, so one can say that more aether energy flows out (in). 
When taking them back to infinity, this energy has to be resupplied as work to readjust the boundary conditions at the surface of the plates: 
the Casimir effect works  as an ideal battery \citep{nieuwenhuizen2023exact}.

While the Casimir energy for parallel conducting plates is negative, it is positive for a conducting spherical shell, which has a tendency to expand 
for small and large radii \citep{boyer1968quantum}. 
While the Casimir energy is always positive,  a bump in the energy as function of the radius exposes
 an intermediate regime with tendency to contract  \citep{cetto1993casimir}.

\subsection{Zero point/aether energy in the black hole interior}

The assumption that {\it aether energy can behave as a fluid} is the corner stone for the present work.
It arose recently as sine qua non element in our singularity-free solution for the black hole interior  \citep{nieuwenhuizen2021interior,nieuwenhuizen2023exact}.
We propose that there is an extended core with a net (positive) charge. In the formation process, 
binding energy released by dissolving the protons and neutrons into free quarks is partly employed for electrostatic energy.
Another part of the binding energy is proposed to be inserted in the quantum aether (without creating particles), 
which acts as a local cosmological constant (LCC).

The Einstein equations impose that the LCC and the electric field coexist.
A change in the charge distribution changes the electric field and imposes a change in the LCC 

\subsection{Zero point energy as a physical entity}
\label{zpe-physical}

In the black hole problem, the zero point energy was treated as physical. Let us motivate this.
As said, the naive (bare) expression for the energy density of the quantum vacuum is divergent.
That a constant can be subtracted from the bare value is compatible with the Callan-Symanzik equation, 
the renormalization equation in quantum field theory \citep{peskin2018introduction}.  
This allows to define the renormalized aether energy density as the physical energy density.
It can be positive, negative or zero.

As for the single harmonic oscillator mentioned above, the aether modes can be deformed
which relates to a finite aether energy density, positive or negative. 
For cosmological applications, no boundaries like Casimir plates or shells are present, but
an electric field due to a mismatch of $+$ and $-$ charges  may take their role.

In the application to the Casimir effect, to BHs  and to the cosmos in the present work, the physical zero point energy does not vanish.
In each of these cases, this is due to a physical effect.
 For the cosmos, we shall assume that ZPE was injected in the aether during the Big Bang, to turn, in the course of time, partly into particles, 
 fields and dark matter.

\subsection{Zero point energy as the dark energy}

As a first step to give zero point energy a more prominent role, we identify the ZPE density of the quantum fields with the sought cosmic dark energy. 
In doing so, we follow our teacher Tini Veltman \citep{veltman1975cosmology} at the University of Utrecht,
who cites Andrei Linde  \citep{linde1974lee} and Joseph Dreitlein  \citep{dreitlein1974broken}, and our colleague Sander Bais  
with coauthor Robert Russell \citep{bais1975magnetic}. The idea  itself goes back to Yakov Zeldovich \citep{zel1967cosmological,zel1968cosmological}.
The ZPE density is non-zero, having the ``renormalized'' (physical) value of 70\% of the critical cosmic mass density $\rho_c\simeq9\,10^{-30}\gr/\cm^3$, 
not the ``bare'' (unphysical) one which is 123 orders of magnitude too large -- but see section \ref{inflation} for a fresh view.

In making this step, we explain -- by default -- the dark energy.
Its merit is that the case is based on known physics and that we can postulate that the sought dark matter originates 
as a specific part of the very same ZPE. 
In section \ref{inflation} we argue that the present small value of the cosmological constant is a dynamical effect, not due to fine-tuning.

\subsection{Electric field as partner in zero point energy condensation}

In the plasma of galaxies and clusters, the first actors are  the free charges. 
The electrons, protons and ions occur at high density $\sim 0.01/\cm^3$, and in principle compensate each other.
Small local mismatches produce an electric field with its negative longitudinal pressure, which has to be compensated by the ZPE
in the way prescribed by the Einstein equations.
Hereto, we view the ZPE as a dynamical quantity: energy stored in the aether  can flow and partly condense on mass concentrations 
such as black holes, galaxies and clusters. 
The free charges are not an accidental property; they provide a skeleton to which the ZPE is attached.

Together, the ZPE and the electric field form a {\it scaffold}, a large correlated structure,  for normal matter.
It  constitutes our {\it proposal for the dark matter}, which gets supported by various arguments.

\renewcommand{\thesection}{\arabic{section}}
\section{Theoretical framework}
\setcounter{equation}{0}%
\renewcommand{\theequation}{3.\arabic{equation}}

\label{Fieldeqs}

\subsection{Full Einstein equations}

 We start in general relativity and express the invariant line element $\d s^2 =g_\mn \d r^\mu \d r^\nu$ 
 in spherical spatial coordinates, $r^\mu=(t,r,\theta,\phi)$ with $\mu=0,1,2,3$, as\footnote{Unless indicated otherwise,  we employ units $\hbar=c=1$, $\eps_0=1/4\pi$, $\mu_0=4\pi$.}  \citep{nieuwenhuizen2021interior}
\BEQ\label{dssqTN}
\d s^2 = \mmin \Sn ^2\bSe \,   \d t^2\pplus \frac{1}\bSe \d r^2\mmin r^2(\d\theta^2\pplus\sin^2 \!\theta\,\d\phi^2), \qquad \bar S=S-1,
\EEQ
 with functions $N(r)$ and $\Se(r)$  to be specified later. As generalization of the Schwarzschild metric, 
 it is the most general spherically symmetric one, see e. g. Weinberg's book \citep{weinberg1972gravitation}.
 It leads to a diagonal Einstein tensor $G^\mu_\ednu$ with elements 
\BEQ &&
G^0_{\ed0}= \frac{S+rS'}{r^2},\qquad  G^1_{\ed1}=\frac{S+rS'}{r^2} +2\frac{N'}{N}\frac{S-1}{r} ,\quad \nn\\&&
G^2_{\ed2}=G^3_{\ed3}=\frac{2S'+rS''}{2r}+\frac{N'}{N}\frac{2S-2+3rS'}{2r}
+\frac{N''}{N}(S-1) ,
\EEQ
where  $G^2_{\ed2}=G^3_{\ed3}$ due to the spherical symmetry.
We express the stress energy tensor $T^\mu_\ednu$  in  parameters $\rho_\lam $ and $\rhoA $ to be specified later, and parameters $\rho_\vth$, $p_{r}^\vth$ and 
$p_\theta^\vth=p_\phi^\vth=p_\perp^\vth$ connected to normal matter ($\vth$), possibly thermal,  with, in principle,  an anisotropic material pressure,
 \BEQ \label{TMnrhomA}
 \hspace{-3mm}
T^\mu_\ednu = \rho_\lambda \delta^\mu_\ednu \pplus \rhoA  \cC^\mu_\ednu  \pplus T^\mu_{\vth\,\nu} ,
\qad  \cC^\mu_\ednu =\diag(1,1,\mmin 1,\mmin 1)  ,
\qad T^\mu_{\vth\,\nu}=\diag(\rho_\vth,\, \mmin p_r^\vth,\, \mmin p_\perp^\vth,\, \mmin p_\perp^\vth)  ,
\EEQ
with all coefficients functions of $r$. They are fixed by the Einstein equations $G^\mu_\ednu = 8\pi GT^\mu_\ednu$
and  may be presented in a mixed way as
\BEQ \label{EMT=}
\label{rhoL=}
&& \hspace{-6mm}
\bar\rho_\lambda
=\frac{2\Se  \pplus 4r \Se  '\pplus r ^2\Se ''}{32\pi G r^2} \pplus 
\frac{3\Sn ' }{\Sn }\frac{2\bar \Se  \pplus  r \Se  '}{32\pi G r} \pplus \frac{\Sn ''}{\Sn }\frac{\bar \Se  }{16\pi G} , 
\\ &&  \hspace{-6mm}
\label{rhoA=}
\bar\rho_E
=\frac{2 \Se  -r^2\Se  ''}{32 \pi  G r^2 }+\frac{\Sn ' }{\Sn }\frac{ 2\bar \Se-3r\Se  ' }{32 \pi  Gr} -\frac{\Sn ''}{\Sn } \frac{\bar \Se  }{16 \pi  G }  ,
\\ &&  \hspace{-6mm}
\label{sigth=}
\bar \rho_\vth =-\frac{\Sn '\bar \Se  }{4\pi Gr\Sn }  , \qquad
\EEQ
where
\BEQ
\bar\rho_\lambda\equiv \rho_\lambda -\frac{p_r^\vth+p_\perp^\vth}{2}  ,\qquad
\bar\rho_E\equiv \rhoA  -\frac{p_r^\vth-p_\perp^\vth}{2} ,\qquad
\bar \rho_\vth \equiv \rho_\vth+p_r^\vth . 
\EEQ
In case the matter has an isotropic pressure, these reduce to
\BEQ
\bar\rho_\lambda\equiv \rho_\lambda - p_\vth  ,\qquad
\bar\rho_E\equiv \rhoA  ,\qquad
\bar \rho_\vth \equiv \rho_\vth+p_\vth . 
\EEQ
When material energy and pressure can be neglected, one simply has $\bar\rho_\lambda=\rho_\lambda$, $\bar\rho_E= \rhoA$ and  $\bar \rho_\vth=0$.
It implies that $N(r)=1$, which simplifies the analysis.

Combining the above expressions,  or by considering the $T^0_{\ed0}$ component, eq. (\ref{rhoL=}) can be replaced by 
\BEQ\label{rhotot2}
\rho_\tot \equiv \rho_\lambda+\rho_E+\rho_\vth=\frac{\Se+r\Se'}{8\pi G r^2}   .
\EEQ
This can be integrated to solve $S(r)$ with vanishing value of $\Se(0)$ as
\BEQ \label{Mtot=}
\Se(r)=\frac{2GM_\tot(r)}{r},\qquad M_\tot(r)=4\pi\int_0^r\d u\, u^2\rho_\tot(u) .
\EEQ
(The singular behavior $\rho_\tot(r)\sim 1/r^2$ for $r\to0$ leads to a finite $\Se(0)$ and a still regular metric.)
Using  (\ref{sigth=}), the function $\Sn$ can be eliminated from (\ref{rhoA=}),  to yield an equation for $\Se$ alone, 
\BEQ
\label{rhoEba=}
\rho_E=\frac{2 \Se  -r^2\Se  ''}{32 \pi  G r^2 }+\frac{r\bar\rho_\vth'}{4}+\frac{p_r^\vth-p_\perp^\vth}{2}
+ \frac{ r\bar\rho_\vth\Se'}{8\bar \Se} - \frac{\pi G r^2\bar\rho_\vth^2}{\bar \Se}  .
\EEQ

The strategy to solve these equations is as follows. The energy densities and pressures of matter and electrostatics are assumed to be known,
so that $\Se$ can be solved from (\ref{rhoEba=}). It will determine $M_\tot$ and $\rho_\tot$ via (\ref{Mtot=}), next $\rho_\lam$ via (\ref{rhotot2});
lastly,  $N$ can be solved from (\ref{sigth=}). 
After these steps, $\bar\rho_\lambda$ can just be read off from (\ref{rhoL=}); being determined by the electric charge density and normal matter, it is ``enslaved''. 

From (\ref{Mtot=}) we can consider the rotation speed for circular orbits, 
\BEQ
v_\rot^2(r)\equiv \frac{GM_\tot(r)}{r} = \half \Se (r),
\EEQ
hence any result for $\Se $ can be expressed in terms of $v_\rot$ or $M_\tot$.

The task simplifies when the energy and pressure (but not the charges) of normal matter can be neglected,
so that $N(r)=1$. For the application to singularity-free, cored black holes, this yields a class of exact solutions \citep{nieuwenhuizen2021interior}.

\subsection{Coulomb electrostatics in electrogravity}

We consider a static potential $A_\mu=\delta^0_\mu A_0(r)$ and define the radial electric field as $E(r)=-A_0'(r)/\Sn (r)$. It equals 
\BEQ
E(r)=\frac{Q(r)}{r^2},\qquad  Q(r)=4\pi\int_0^r\d u\, u^2 \rho_q(u).
\EEQ
Here $Q(r)$ is the total charge enclosed within radius $r$, given the charge density $\rho_q$.
Since  the metric (\ref{dssqTN}) is diagonal, the  stress energy tensor keeps its special relativistic form,
\BEQ
T^\mu_{E\,\nu}=\rho_E\cC^\mu_\ednu,\qquad \rho_E(r)=\frac{E^2(r)}{2\mu_0}=\frac{E^2(r)}{8\pi},
\qquad \cC^\mu_\ednu =\diag(1,1,\mmin 1,\mmin 1) ,
\EEQ
as employed in (\ref{TMnrhomA}).
$T_E$ is traceless, and involves a positive energy density and transversal pressures, but a negative longitudinal presssure;
all equal in magnitude.

\subsection{Linearized Einstein equations}
\label{linEineq}

For application to galaxies and clusters we can linearize around the vacuum $\Sn =1$ and $\Se =0$.
Indeed, the observed rotation speeds $v_\rot$ are much smaller than the speed of light $c=1$, so that $\Se(r)=2v_{\rm rot}^2(r)\ll 1$.
 Eq. (\ref{sigth=})  yields
 \BEQ
 \label{Npr=}
N(r)=1- 4 \pi G \int_r^\infty \d u\, u [\rho_\vth(u)+p_\vth(u)].
\EEQ
In linearized form, Eq. (\ref{rhoEba=}) reads
\BEQ
\label{rhoEbad=}
\frac{2 \Se  -r^2\Se  ''}{32 \pi  G r^2 }= \rho_E-\frac{r(\rho_\vth'+p_\vth')}{4}.
\EEQ
The homogenous solutions are $\Se=r^2$ and $1/r$. The proper inhomogeneous solution,
\BEQ 
 \label{Mtotr=} \hspace{-5mm}
M_\tot(r)=\frac{r\Se (r)}{2G}=M_\Lam+M_\vth^\rho+M_\vth^p+\frac{4}{3}M_E
+\frac{4\pi}{3}r^3\rho_E^>,\quad \rho_E^>(r) \equiv 4\int_r^\infty \d u\frac{\rho_E(u)}{u} ,
\EEQ
obeys $M_\tot(0)=0$ and $S(r)\sim r^2$ at small $r$. The involved functions are
\BEQ \label{MErhop=}
(M_E,M_\vth^\rho,M_\vth^p)=4\pi\int_0^r\d u\,u^2(\rho_E,\rho_\vth,p_\vth) ,\quad M_\Lambda=\frac{4\pi}{3} \rho_\Lambda r^3 =\frac{\Lam r^3}{6G}.
\EEQ
where $\Lambda$, the standard cosmological constant, enters as an integration constant.
Surprisingly, the last term in (\refl{Mtotr=}) involves the outer region $u\ge r$. (This would be avoided by subtracting its integral from 0 to $\infty$, 
which connects to a homogenous solution $S\sim r^2$, but, acting at large $r$ as a huge negative cosmological constant, that is not acceptable.)

Equating (\ref{Mtotr=}) to the integral of $\rho_\tot$ in (\ref {rhotot2}), viz. $M_\tot=M_\lam+M_E+M_\vth^\rho$,
we can identify the LCC component,
\BEQ \label{MLam=}\label{Mlam=}
M_\lambda(r)=M_\Lam(r) +M_\vth^p(r)+ \frac{1}{3}M_E(r) +\frac{4\pi}{3}r^3\rho_E^>(r).
\EEQ
The total excess ZPE $(M_\lam- M_\Lam) \vert_{r\to\infty}=\frac{1}{3}M_E(\infty)+M_\vth^p(\infty)$ is assumed to have flown in from infinity
in the dynamics towards this phase. Notice that even without electric field, the aether already supplies  the pressure term
$M_\vth^p$, often called a ``relativistic correction'' to the matter term $M_\vth^\rho$.  

The derivatives $M_\tot$ and $M_\lam$ yield
\BEQ &&
\label{roLa=}
 \rho_\tot=\rho_\Lam +\rho_E^> +\rho_\vth+p_\vth, 
  \qquad 
  \label{rhol=}
 \rho_\lambda=\rho_\Lam+\rho_E^> -\rho_E +p_\vth.
 \EEQ

Setting $T^\mu_\ednu = {\rm diag}(\rho_\tot,-p_\tot^r,-p_\tot^\theta,-p_\tot^\phi)$,  this collects
\BEQ \hspace{-3mm}
\rho_\tot=\rho_\lam+\rho_E+\rho_\vth,\quad 
p_\tot^r=-\rho_\lam-\rho_E+p_\vth ,\quad
\label{prptpp}
p_\tot^\theta=p_\tot^\phi=p_\tot^\perp=-\rho_\lam+\rho_E+p_\vth.
\EEQ
The total pressure is always anisotropic.
After eliminating $\rho_\lambda$, the pressures (\ref{prptpp}) read
\BEQ \label{pitot}
p_\tot^r=-\rho_\Lam-\rho_E^> 
,\qquad  
p_\tot^\perp=-\rho_\Lam - \rho_E^> + 2\rho_E.
\EEQ
The cosmological constant $\rho_\Lam$ has been included for completeness;  for applications to galaxies and clusters it is negligible.
Outside a region with matter and a net charge $Q$, $\rho_E=\rho_E^>=Q^2/8\pi r^4$ leads to $\rho_\lam =\rho_\Lam+p_\rmm$.
For galaxies and clusters, the net included charge will vanish at some radius, which can be identified with their outer border.

\subsection{Electro-zero-point  energy (electro-aether energy) as the dark matter}

We have now arrived at the point to introduce \EAE   as the combination of the electric field energy and the ZPE in the LCC. 
In this paradigm we can identify the DM parts,
\BEQ \label{rhodm=}
&& \hspace{-6mm}
\rho_\dm=\rho_\lam+\rho_E-p_\vth=\rho_E^>,
\quad M_\dm=\frac{4}{3}M_E+\frac{4\pi}{3}r^3\rho_\dm,
\\&& \hspace{-6mm} \label{pdm=}
 p_\dm^r = -\rho_E^> , 
 \qquad
 p_\dm^\perp=2\rho_E-\rho_E^>, \qquad \rho_E^>=4\int_r^\infty \d u\frac{\rho_E(u)}{u},\qquad \rho_E=\frac{E^2}{8\pi} .
 \EEQ
For small $r$, the expression for $\rho_\dm$ typically leads to a constant value, that is to say,  
\EAE   naturally involves constant-density dark matter cores. For empirical support of  this, see sec. \ref {sec:galcore}.
Nevertheless, powerlaw behavior $S\sim r^{2n}$ at small $r$ with $n\ge0$ is possible, with $\rho_\lam\sim\rho_E\sim r^{2(n-1)}$ and
$E\sim r^{n-1}$, $\rho_q\sim r^{n-2}$. The extreme case is $n=0$, where $S(0)$ is a constant between 0 and 1.

The build up of a net positive charge inside a central region of a galaxy or galaxy cluster, implies that some electrons must move outwards.
Let us introduce a crossover radius by $R_{\rm co}$; beyond it, the expelled electrons make up a net negative charge density.
The included net charge $Q(r)$ grows up to $R_\co$ but decays beyond it, making $\rho_E=Q^2/8\pi r^4$ decay quicker than $1/r^4$ at
moderately large $r$.  The radial {\it size} of the {\it galaxy (cluster)}  $R_g$ can be defined as the radius where $Q(r)\to0$; 
due to the large strength of electrostatics, this is a {\it sharp} boundary.

Approximating $v^2(r)$ here by $GM_\dm(r)/r$ (dark matter only) and denoting $v^2(r)=v_2(t)$ and $Q^2(r)=r^2q_2(t)/4G$ with $t=\log(r/$kpc), it follows that
\BEQ\label{q2dot}
2q_2+\dot q_2=4v_2+4\dot v_2-\ddot v_2-\dddot {v}_{\hspace{-0.95mm}2}.
\EEQ
The crossover radius where $Q'(R_\co)=0$, now set by the condition $2q_2+\dot q_2=0$, is related to the rotation curve.
In integral form Eq. (\ref{q2dot})  reads
\BEQ
q_2=2v_2+\dot v_2-\ddot v_2.
\EEQ
These relations, extended to include normal matter, offer a test between the electric charge (and field) profile and the rotation curve.

\subsection{A toy galaxy}
\label{sec:toygal}

\begin{figure}\label{figrhozpRiRe} 
\begin{center}
\includegraphics[width=8cm]{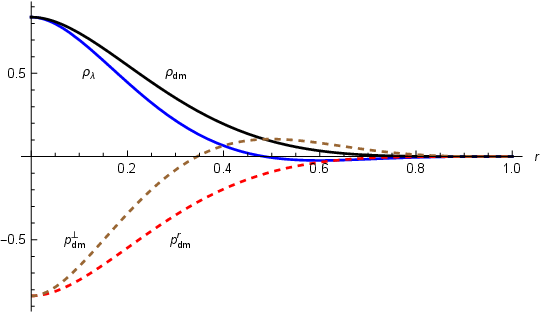}
 \end{center}
\caption{
The local cosmological constant $\rho_\lam$ starts with positive value but has a negative tail, 
as shown here for a toy galaxy with charge distribution (\ref{toyrhoq}).
While $\rho_\dm$ is strictly positive and $p_\dm^r$ strictly negative, 
$p_\dm^\perp$ starts out negatively and has a positive tail.
 }\label{figeen}
\end{figure}

Let us consider a toy galaxy, upon neglecting $\rho_\Lam$, $\rho_\vth$  and $p_\vth$. 
It has normalized radius  $R_g=1$ and normalized net charge density,
\BEQ\label{toyrhoq}
\rho_q(r)=(1-r)(3-5r) ,\quad (0<r<1)  ; \qquad \rho_q=0,\quad (r>1),
\EEQ
which is positive for $0<r< R_\co=\frac{3}{5}$ and negative up to $R_g$. The included electric charge and energy density are
\BEQ
Q(r)=4\pi r^3(1-r)^2,\quad \rho_E=2\pi r^2(1-r)^4 ,
\EEQ
with $Q(R_\co)=432 \pi /3125=0.4343$. The total charge vanishes,  $Q(1)=0$.
Eq.(\ref{rhodm=}) yields
\BEQ \label{profiles}
&& \rho_\lam=\frac{2\pi}{15}(1-r)^4(2+8r-25r^2),\qquad  \rho_\dm=\frac{4\pi}{15}(1-r)^5(1+5r), \nn\\&&
p_\dm^r=-\frac{4\pi}{15}(1-r)^5(1+5r),\qquad \quad p_\dm^\perp=-\frac{4\pi}{15}(1-r)^4(1+4r-20r^2) .
\EEQ 
While $\rho_\dm$ is strictly positive and $p_\dm^r$ strictly negative, $\rho_\lam$ is positive up to  $r_0=0.4849$
and then negative beyond. Notice that $r_0<R_\co=0.6$. 
Conversely, $p_\dm^\perp$ is negative up to  $r=$ $0.3449$ and positive beyond. 
 The profiles (\ref{profiles}) vanish at $R_g=1$. They are plotted in fig. \ref{figeen}.
 That $\rho_\lam$ has generally a negative tail follows from eq. (\ref{rhol=}) near $R_g$, where $\rho_E$ vanishes.

The accumulated  zero point energy,
\BEQ
M_\lam(r)=\frac{8\pi^2}{15}\left (\frac{2 r^3}{3} -9 r^5 +\frac{70 r^6}{3} -\frac{180 r^7}{7} +\frac{27 r^8}{2} -\frac{25 r^9}{9} \right) ,
\EEQ
has total $M_\lam\equiv M_\lam(1)=4\pi^2/945=0.04177$. 
Its positive density part between 0 and $r_0$ contains $M_\lam(r_0)=1.652M_\lam$,
while the tail between $r_0$ and 1 has value $-0.652M_\lam$.
With respect to  $M_\lam$, the ZPE  imported from remote surroundings, this exhibits an additional  $65.2\%$ 
taken out of the vacuum in the outer  region  $r_0<r<1$, and transferred to the inner region  $r<r_0$.
With vanishing net total charge, $Q(r)=0$ for $r\ge R_g$, the total electrostatic energy $M_E=4\pi \int_0^1\d r\,r^2\rho_E(r)$ equals $3 M_\lam$.

These properties reflect our main assumption: zero point (vacuum) energy can be taken out or put into the (quantum) aether, 
at amounts governed by the Einstein equations.

\subsection{The matter components}

In a galaxy there is normal matter such as stars, planets, free floating planets, hydrogen gas clouds, and a plasma of protons, ions and electrons.

In a galaxy cluster, of all the galaxies, only the brightest cluster galaxy (bcg), located in the center and often the brightest X-ray source in the cluster, 
brings a relevant contribution to the enclosed mass $M_\tot(r)$; it can be described by its mass density $\rho_\bcg(r)$. 
This galaxy will also contain dark matter, which is a part of $\rho_\dm$.
Our fit profile $\rho_\bcg$ in section \ref{IsoLensingfit} combines normal and dark matter; as of now, there are no data for the separate parts.
For simplicity of notation, we shall nevertheless write 
\BEQ\label{rhompm=}
\rho_\vth=\rho_\bcg +\rho_g,\qquad p_\vth=p_g,
\EEQ
 where $\rho_g$ and $p_g$ describe the X-ray gas, see eq. (\ref{rpgas}) below.

\subsection{The X-ray gas }

Free electrons, accompanied by protons and ions, occur in galaxies and clusters. 
The mass density $\rho_g=n_e\bar m_N$ involves the local free electron density $n_e$ and the average molecular weight $\bar m_N\approx (7/6)m_N$ 
with $m_N$ the nuclear weight \citep{morandi2010unveiling}. The pressure is $p_g=n_eT$, where the typical scale of $T(r)$ is several keV.
So the X-ray gas involves
\BEQ \label{rpgas}
\rho_g(r)=n_e(r)\bar m_N,\quad  p_g(r)=n_e(r)T(r).
\EEQ
With $T$ in the keV regime, the gas is hot but the protons and nuclei are nonrelativistic, with $p_g/\rho_g\sim T/m_N\lesssim 0.01$.
 
 In the \EAE   approach the X-ray plasma is indispensable, since by setting up a small charge mismatch (estimated in sec. \refl{charge-mism}),
 it creates the electric field that combines with ZPE as ``dark matter''.
 Turned around, in this paradigm the presence of DM requires the presence of a net charge density and hence a (partial or full) plasma.

 \subsection{Hydrostatic equilibrium}
 \label{sec:hydeq-1}

For a fluid element in equilibrium, the balance of forces is expressed by hydrostatic equilibrium.
Expressed by the energy conservation $T^\mu_{\ednu;\mu}=0$, this
is automatically satisfied for a bona fide solution of the Einstein equations, since  $G^\mu_{\ednu;\mu}=0$ by construction.
For the stress energy tensor (\ref{TMnrhomA}) in the metric (\ref{dssqTN}) this leads to the exact  force balance
\BEQ \label{new}
p_\vth'+ p_\lam'=  \cFE+\cFG  \quad \text{or}\quad \cF_\vth+\cF_\lam+\cFE+\cFG=0, \quad
\EEQ
where we employed  the cosmological pressure $p_\lam=-\rho_\lam$.
In exact form, 
the respective force densities read, using $\rho_E=E^2/8\pi$, $E=Q(r)/r^2$ and $Q'=4\pi r^2\rho_q(r)$,
\BEQ \label{TOVtheo}
&&  \hspace{-1cm}
 \cF_\vth=-p_\vth',\qquad
 \cF_\lam=-p_\lam'=\rho_\lam' , \qquad 
\cFE=\rho_E'+4\frac{\rho_E}{r} =\frac{Q'Q}{4\pi r^4} =\rho_q(r)E(r),
\nn \\ && \label{newlam2}
\hspace{-1cm}
\cFG =-\bar\rho_\rmm\left(\frac{S'}{2\bar S}+\frac{N'}{N}\right) =
-\frac{G}{r^2}\left(\rho_\rmm+p_\rmm\right)\frac{M_\tot-4 \pi  r^3(\rho_\lam+\rho_E-p_\vth)}{1-2GM_\tot/r} .
\EEQ
Here $\cF_E=\rho_qE$ is the Coulomb force on a unit volume with charge density $\rho_q$ in an electric field $E$.
Eq (\ref{rhol=}) allows to replace the combination $\rho_\lam+\rho_E-p_\vth$ in $\cF_G $ by $\rho_\Lam+\rho_E^>$.

In \EAEc, the leading terms are $\cF_\lam$ and $\cFE $; neglecting $\cFG $, one recognizes the derivative of $\rho_\lam$ in eq. (\ref{rhol=}) 
as the solution for hydrostatic equilibrium.  The exact approach adds the nonlinear, Newtonian $\cFG $ term, which is small, see sec. \ref{hydro-stab}.
Integrating (\ref{new}) from $r$ to $\infty$ yields  the full nonlinear correction to $\rho_\lam$; it becomes a self-consistent relation, since
$\rho_\lam$ enters $\rho_\tot$ through eq. (\ref{roLa=}), and hence through $M_\tot$. 

The fact that $\cFG $ contributes  to $\rho_\lam'$, exhibits the malleability (enslavement) of the ZPE/AE, doing just the right thing in the situation at hand.

 \renewcommand{\thesection}{\arabic{section}}
\section{Physical estimates}
\setcounter{equation}{0}
\renewcommand{\theequation}{4.\arabic{equation}}

\label{phys-estimates}

\subsection{Charge mismatch in the plasma}
\label{charge-mism}

For the rotation speed $v^2=GM/r$, eq. (\ref{Mtotr=}) estimates the electric field as $E\sim v/c\ell_Pr$.
In the galaxy clusters A1689 and A1835, the free electron density in the center is of order $n_e(0)= 0.05-0.09$/cm$^3$ \citep{nieuwenhuizen2021accurate}.
The density $n_p$ of plus charges (protons, ions) is very close to this.
Equating $E$ to $e(n_p-n_e)r$ yields the relative mismatch $\del_q\equiv (n_p-n_e)/(n_p+n_e)\sim v/cen_e\ell_Pr^2
\sim 6\,10^{-15}$ for $v=1000$ km/s and typical scale $r=100$ kpc. 

In our Galaxy near the Sun $n_e\sim 0.02$/cm$^3$ leads for $r=10$ kpc and $v=200$ km/s to  $\del_q\sim 10^{-13}$.  
In 2013, the Voyager 1 spacecraft observed electron plasma oscillations corresponding to an electron density $n_e=0.08/\cm^3$, 
very close to the value expected in the interstellar medium and confirming the order of magnitude \citep{gurnett2013situ}.
The measured electric field of about 1 $\mu$V/m at kHz frequencies does not refer to a static field.

These tiny mismatches express that the plasma is not completely neutral, but slightly charged.
The electric field is an average of spatially and temporally 
fluctuating $E$ fields with their related $B$ fields caused by the relatively high density of moving protons, ions and electrons.
Since they are so light,  it is easier to push out electrons than protons and ions; hence we assume that the inner galaxy is positively charged,
and the outskirts negatively, separated by a crossover radius $R_\co$.
Due to the strength of the Coulomb force, a galaxy has has a sharp boundary $R_b$, where the included charges cancel each other.

In reverse, the locally available amount of ZPE 
sets the size of the static  $E$ field. Such a correlation can be searched for via the determination of the dark matter, also in the intragalactic medium.

\subsection{Galactic electric field}

The possibility of a galactic electric field has been considered in the literature.
Ref. \citep{reucroft2014galactic}  estimates from rotation curves a field of 1 V/m at the solar location. 
To keep most electrons in the stellar interior, ref. \citep{chakraborty2014possible} estimates
a center-to-surface potential difference of 1000 V; giant galaxies to have a similar potential difference, and rich clusters $\sim$10 kV.

In our approach, the electric field joins the ZPE to make up the \EAEd. The DM density in the solar neighborhood is
$\rho_\dm^\odot = 0.35^{+0.08}_{-0.07}$ GeV/cm$^3=6.2^{+1.4}_{-1.2}10^{-25}\, \gr/\cm^3$  \citep{kafle2014shoulders}.
 A recent, sharper estimate is 
$\rho_\dm^\odot = 0.447^{+0.004}_{-0.004}$ GeV/cm$^3=7.96^{+0.07}_{-0.07}10^{-25}\, \gr/\cm^3$ \citep{ou2024dark}.
The total mass density  $ (0.097\pm 0.013 )M_\odot/ {\rm pc}^3=(6.6\pm 0. 9)10^{-24}\gr/\cm^3$ is about 8 times larger.
From eq. (\ref{rhodm=}) we estimate that maximally
\BEQ 
\rho_\dm^\odot=4\int_r^\infty\d u\frac{\rho_E(u)}{u}\approx 4\rho_E^\odot\log\frac{R_c}{r}\approx 10\rho_E^\odot, \qad   
\EEQ
for $R_c=100$ kpc and $r=8.1$ kpc. Thus taking 10\% of the DM density as electrostatic energy $E^2/8\pi$ leads to a local field 
$E_\odot=0.042\, \statV/\cm=1.3$ kV/m. 
It is likely shielded by the ionosphere.
Notice that the Earth electric field is about 0.1 kV/m under fair weather conditions  \citep{rycroft2000global}.

In sec.  \ref{HEcluster-EVE} and  figure \ref{figstaticE} 
we predict similar electric fields of $\sim 1.5$ kV/m and $0.5$ kV/m in the central $\sim 500$ kpc of the galaxy clusters A1689 and A1835, respectively.

\subsection{Seeding of galactic magnetic fields}

Large scale, quasi static magnetic fields abound in the cosmos and play an important role in the evolution of galaxies, 
but their origin is still debated \citep{beck2013large}.
In \EAE   theory, the formation of a dark matter core is a dynamical effect with a slowly varying electric field, which, according to Maxwell law
$\eps_0 \dot\vE=\nabla\times\vH$, generates cosmic magnetic fields. 
In the local Galaxy and in the fat clusters A1689 and A1835 we predict electric fields in the 1 kV/m range, 
which corresponds in strength to a magnetic field of 0.029979 G, while observed magnetic fields lie typically in  the 1-10 $\mu$G regime.

As to orders of magnitude, Maxwell's law yields the estimate $B\sim LE/ct$.
For the Galaxy, $E\sim 1$ kV/m, $L\sim 100$ kpc and $t\sim10$ Gyr imply $B\sim 1\mu$G, and for clusters with $L\sim 1$ Mpc it yields $B\sim 10\mu$G.
As these rough estimates produce the right order of magnitude, this connection deserves a detailed analysis.

\subsection{Implementation of hydrostatic stability}
\label{hydro-stab}

The net-charge ratio $\delta_q\sim 10^{-13}-10^{-14}$ needed in the \EAE   approach is actually relatively large and unexpected. 
 A standard argument is to consider a sphere with mass $M=Nm_N$ and charge $Q=\del_qNe$. 
In principle, the ratio of Coulomb and Newton forces at the surface, $Q^2/GM^2=(Qm_P/M)^2$, can not exceed unity, 
which occurs for $\del_q=m_N/2em_P=0.5\, 10^{-18}$,
quite smaller than the above estimates. Including dark matter by setting $M\approx  2Nm_N$ only alleviates this to $10^{-18}$.
While a huge charged cosmic object seems unrealistic, in the standard Einstein-Maxwell theory there cannot be a stable shell with 
negative charge density around a positively charged core, as it would add to rather than compensate the Newton attraction.
Nevertheless, the typical value of only $\delta_q\sim 10^{-18}$ looks problematic for \EAE   theory, so let us analyze the situation.

In the \EAE   setting, the galaxy or cluster has a large core with net positive charge density, surrounded by a halo of negative net charge density. 
As a whole, it is not charged.
One may define the boundary of a galaxy or cluster $R_b$ as the radius where $Q(R_b)=0$.

In the interior $r<R_b$,  our starting point is eq. (\ref{new}), the hydrostatic equilibrium condition for the balance of forces acting 
on a fluid element. In terms of the local cosmological pressure $p_\lam=-\rho_\lam$, this takes to leading order the form
 \BEQ \label{HE13}
 p_\rmm'+p_\lam' =\cF_E+\cFG , \quad \cF_E =  \rho_E'+4\frac{\rho_E}{r}=E\rho_q,\quad 
 \cFG \approx -G (\rho_\rmm+p_\rmm)\frac{M_\tot}{r^2} .
\EEQ
In absence of $\rho_E$ and $p_\lam'$ as in $\Lam$CDM, the pressure must decay in such a way that $\cFG $, the Newton attraction per unit volume, is balanced.
But notice that, according to (\ref{rhompm=}),  $\rho_\vth=\rho_\bcg +\rho_g$ includes both the gas and the normal matter mass densities. 
As we discuss in section \ref{HEpuzzle},  this approach does not work out properly in clusters.

In \EAEc, the finite $\cF_E=E\rho_q$ is the local Coulomb force density,  involving a positive $E=Q(r)/r^2$ and a  $\rho_q$  that is positive for $r<R_\co$ 
and negative for $r>R_\co$.

We can estimate $\cF_G/\cF_E$ as $(\rho_\vth/\rho_E) (GM_\tot/r)\lesssim (GM_\tot/r)\sim 10^{-7}$ for a galaxy with $M_\tot=10^{11}M_\odot$ and $r=100$ kpc,
and $\sim5\,10^{-5}$ for a fat cluster with $M_\tot=10^{15}M_\odot$ and $r=1$ Mpc.
These values play the role of the above factor $(m_N/\del_qem_P)^2$, showing that 
$\delta_q$ is relatively large, and $\cFG$ relatively small,  in the presence of \EAEd.
Clearly, in eq. (\ref{HE13}) the combination $p_\rmm'+p_\lam'$ should be balanced, not $p_\vth'$ alone.
Because $p_\vth'\sim\cFG $ can even be neglected,  the balance is in essence $p_\lam'=\cFE$,
 obeyed grace to eq. (\ref{rhol=}).

In conclusion, the above Newtonian argument fails, since it overlooks the stress (pressure) related to a positive (negative) aether energy density.
On a fluid element of \EAEc, the strong Coulomb force (repulsive for $r<R_\co$ and attractive for $r>R_\co$) is balanced by an equally strong 
(inward c.q. outward) force from the zero-point/aether pressure gradient. 
 As a result, galaxies and clusters with charge ratios $\del_q$ well exceeding $10^{-18}$ are allowed in \EAEd.

\subsection{Metastability vs instability}
\label{metastab-instab}

Hydrostatic equilibrium for patches of matter describes stability at the macroscopic scale, but it does not automatically imply equilibrium at the microscopic scale. 
Indeed, an individual ion is affected by the strong outward Coulomb force but not know to involve a restoring force from the cosmological pressure, 
and neither an effect of the outer shell of negative charges. Likewise, an electron is strongly attracted inwards,
the ones moving from $r>R_\co$ to $r<R_\co$ will lessen the charge mismatch. 
While for a reshuffling of the charge distribution, an accompanying reshuffling of the ZPE is required, an inherently unstable situation may remain.

This observation suggests that \EAE   cores are unstable, which in principle demonstrates the failure of the theory.
But the involved timescales need not coincide, and may be cosmological, so that metastable cores on Gyr timescales are compatible with \EAEd.

We estimate the effects by connecting to observations. While galaxies and clusters clearly exist with dark matter supposedly arising from \EAE   theory,
there are also indications for their subsequent dissolution.
For galaxies these are discussed in sec.  \ref{reverse instab}.

 \renewcommand{\thesection}{\arabic{section}}
\section{Electro-aether-energy in black holes}
\setcounter{equation}{0}
\renewcommand{\theequation}{5.\arabic{equation}}

\label{secBHs}

\subsection{Black hole metrics with a macroscopic core}

The present work was inspired by our solutions for black holes (BHs) with a smooth core enclosed by the inner horizon and an empty 
mantle up to the event horizon.
Assuming that in the stellar collapse a bit more electrons than protons were ejected, the core-collapsed BH will be positively charged.
The protons may dissolve into up and down quarks, thereby releasing their binding energy, 98\% of their mass.  
Upon neglecting the quark and electron masses, analytical solutions were presented based on what we now call \EAE   theory. 
The stress energy tensor takes the form
\BEQ \label{TMnexactBH}
T^\mu_\ednu(r)=\rho_\lam(r)\delta^\mu_\ednu+\rho_E(r)\cC^\mu_\ednu . 
\EEQ
Given the charge distribution and hence $\rho_E$, a class of exact solutions of  the Einstein equation (\ref{rhoA=}), viz. 
$2 \Se -r^2\Se '' = 32 \pi  G r^2\rho_E$, was presented, after which the LCC (ZPE density)
$\rho_\lam$ follows from (\ref {EMT=}), $\rho_\lambda =(2\Se  \pplus 4r \Se  '\pplus r ^2\Se '')/32\pi G r^2$, 
 under the condition that it matches the trivial aether $\rho_\lam=0$ at the inner horizon.
 Special cases were worked out, and  it was found that always $M_\lam=\frac{1}{4}M$, $M_E=3M_\lam=\frac{3}{4}M$.
Next, accounting for quark and electron masses was carried out in a numerical analysis.
 It was found that the fluctuation spectrum has oscillating modes, but no growing (unstable) ones. 

Notice that with $\rho_\tot=-p_\tot^r=\rho_\lam+\rho_E$
the the first law of thermodynamics is locally satisfied in the form $\d U(r)\equiv \rho_\tot(r) \d V= -p_\tot^r(r)\d V$
with $\d V=4\pi r^2\d r$, confirming that neither a temperature nor a chemical potential is connected to \EAEd.

\subsection{Super-Eddington accretion by aether energy inflow}

Black holes grow by mass accretion. It falls firstly on the accretion disk, and from there into the back hole or get spit out in the jets. 
Elsewhere, we show that black holes can also become more massive by inflow of aether energy. Here neither an accretion disk nor
angular momentum plays a role. 
But there is a caveat. In the limit where the energy density (but not the charge density)
of the normal matter can be neglected, we have presented a class of exact solutions. They involve a charge distribution with a net charge
with charge-to-mass ratio $Q/M\sqrt{G}\ge \frac{1}{2}\sqrt{3}$. When aether energy is absorbed, without additional charges, this limit
will be reached. 
Further mass increase needs charge increase, which can come from the surroundings.
This hints at a common growth of the supermassive black hole and the galaxy.
Such has been detected and known as the $M-\sigma$ relation.

The largest known black hole mass,  $66\, 10^9M_\odot$ for Ton 618,  is likely determined by super-Eddington accretion assisted by 
or even dominated by \EAE   absorption. 

Recent observations support BH growth related to ZPE/AE . 
Ref. \citep{farrah2023preferential} reports statistics on  hundreds of elliptical galaxies in 3 redshift bins in the domain $0 < z \lesssim 2.5$,
 showing that the supermassive BHs in massive, red-sequence elliptical galaxies must have grown in mass 
by a factor of 7 from $z \sim 1$ to $z = 0$,  and a factor of 20 from $z \sim 2$ to $z = 0$.

GN-z11  is an exceptionally luminous  galaxy at redshift $z=10.6$. Its spectral features indicate that GN-z11 hosts an accreting black hole.
The assumption of local virial relations leads to a black hole mass of $\log_{10} (M_{BH}/M_\odot) = 6.2 \pm 0.3$, 
which amounts to accretion at about 5 times the Eddington rate \citep{maiolino2024small}. 

Both the existence  of early black holes, even primordial ones (see later sections), 
 with their super-Eddington accretion,  is a central tenet of \EAE   theory.

\subsection{The final parsec problem under \EAE   accretion}
\label{final-parsec}

For two widely separated black holes to become bound and finally merge, potential energy must be lost. 
This can occur by dynamical friction, whereby kinetic energy is transferred to nearby matter.  
For example, a bypassing star can get a slingshot in which it gains kinetic energy
and the BH pair becomes more bound.  When the pair has a separation of a few parsecs,
 there is not enough matter to effectively continue this process, while gravitational radiation becomes relevant at distances of 0.001--0.01 pc.
This is called {\it the final parsec problem} \citep{milosavljevic2003final}. Various ways out have been proposed, including merging 
with help of further stars or a third BH.
Also disk accretion during the merger of supermassive black hole binaries in galactic nuclei works for them \citep{armitage2002accretion}.

In the \EAE   paradigm, merging happens in a galaxy with a  dark-matter core, which gets continuously filled up
while being depleted by the BHs. For a BH pair this feeding increases both masses and also diminishes their distance,
as we now discuss by a standard analysis.

For a two body problem, like a BH pair, the kinetic and potential energies are
\BEQ
K=\half m_1\dot\vr_1^2+\half m_2\dot\vr_2^2, \qquad V=-G\frac{m_1m_2}{r} .
\EEQ
 When $m_{1,2}$ depend on time,  this reads in terms of the mutual position $\vr=\vr_1-\vr_2$  and the barycentre 
 $\vR=(m_1\vr_1+m_2\vr_2)/(m_1+m_2)$ as
\BEQ
K=\half M\left(\dot\vR-\frac{\dot m_1m_2-m_1\dot m_2}{(m_1+m_2)^2}\vr \right)^2+\half\mu\dot\vr^2,\quad 
\EEQ
with, as usual, the  total and reduced masses
\BEQ
M=m_1+m_2,\quad \mu=\frac{m_1m_2}{m_1+m_2} .
\EEQ
The center of mass motion involves the conserved momentum
\BEQ
\vP(t)\equiv M\left(\dot\vR-\frac{m_2\dot m_1-m_1\dot m_2}{(m_1+m_2)^2}\vr \right)=\vP_i ,
\EEQ
so that the centre-of-mass energy $\vP_i^2/2M$ decreases on mass inccrease.
In the frame where $\vP_i={\bf 0}$,  the energy reads as for constant masses,
\BEQ
E=K+V
=\half\mu\dot\vr^2-G\frac{m_1m_2}{r} =\frac{1}{2\mu}\vp^2 -G\frac{\mu M}{r} , \qquad \vp=\mu\dot\vr, 
\EEQ
which leads to a $\dot\mu$ term in the equation of motion
\BEQ\label{t-orbit}
\frac{\d}{\d t}\vp=\mu\ddot\vr+\dot\mu\dot\vr=-G\frac{\mu M}{r^3}\vr , 
\EEQ
For the time-dependent masses, it results in the rate of change of energy
\BEQ
\dot E
=-\frac{\dot\mu}{\mu}K+\left(\frac{\dot \mu}{\mu}+\frac{\dot M}{M}\right)V ,
\EEQ
with a minus sign in the kinetic term.
As expected, only derivatives of the masses, not of the orbit, occur. 
The angular momentum in the rest frame
\BEQ \label{L=Li}
\vL(t)=\vr\times\vp=\mu\,\vr\times\dot\vr=\vL_i ,
\EEQ
is conserved due to (\ref{t-orbit}). 

If $m_{1,2}$ change negligibly during one period, circular orbits involve $K= -E$, $V= 2 E$,
leading to $\dot E/E=3\dot\mu/\mu+2\dot M/M$.
Integration from $t_i$ to $t_\itf$ and $E=-G\mu M/2r$ yields
\BEQ\label{EfEi}
 \frac{E_\itf}{E_i}= \frac{\mu_f^3M_f^2}{\mu_i^2M_i^2}
 =\frac{\mu_\itf m_{1\itf}^2m_{2\itf}^2}{\mu_i m_{1i}^2m_{2i}^2} , \qquad\quad
 r_\itf =\frac{\mu_i m_{1i}m_{2i}}{\mu_\itf m_{1\itf}m_{2\itf}} r_i ,
\EEQ
with $i$ denoting initial and $f$ final values, showing tighter binding upon mass accretion, $r_\itf\sim 1/m_\itf^3$ . 
For Kepler orbits,  the relations for $E_f/E_i$ holds due to the virial theorem $\langle K\rangle= -E$, $\langle V\rangle= 2 E$.
The relation between $r_\itf$ and $r_i$  still holds for the epicentre and the apocentre.

{{The Lagrange-Runge-Lenz ellipticity vector  $\veps$
and ellipticity parameter $\eps$, with $0\le\eps\le 1$, read
\BEQ
\veps=\frac{\dot\vr\times\vL}{G\mu M }-\hat\vr, \qquad 
\eps=|\veps|=\Big(1+\frac{2EL^2}{G^2 \mu^3M^2}\Big)^{1/2}
\EEQ
From (\ref{L=Li}) and (\ref{EfEi}) it is seen that the ellipticity is essentially constant for slow mass increase.}}

For $m_1=m_2=m$ one has $ \mu=m/2$.
Merging can be estimated to happen at the Schwartzschild value $r_\itf=2Gm_\itf$. This leads to a final mass $m_\itf=m_i(r_i/2Gm_i)^{1/4}$. 
In terms of  $m_i=\bar m_i\,M_\odot$ and $r_i=\bar r_i$ kpc, this reads $m_\itf =10^4 (\bar m_i^3\bar r_i)^{1/4}M_\odot$,
an appreciable mass gain from the aether. Relativistic corrections and  gravitational radiation may bring corrections of order unity.

\renewcommand{\thesection}{\arabic{section}}
\section{Electro-aether-energy  in galaxies }
\setcounter{equation}{0}
\renewcommand{\theequation}{6.\arabic{equation}}

\label{EVE-gals}

\subsection{Relation to MOND} 
\label{sec5.1}

Observations by Vera Rubin and Kent Ford in the seventies demonstrated that in the outer part of galaxies, circular orbits have 
nearly the same rotation speed \citep{rubin1970rotation}, constituting ``flat rotation curves". To explain this,
Modified Newtonian Dynamics (MOND) was introduced by Milgrom in 1983 \citep{milgrom1983modification}.  It assumes that the Newton force
decays as $1/r$ at large $r$, which can be expressed as an effective enclosed mass behaving as $M(r)\sim r$. 
For recent reviews, see  \citep{lelli2017one,banik2022galactic}.

For a proper amount of \EAEc, our approach can explain the MOND results.
Let the $\rho_E$ profile take the form of an isothermal sphere, $E^2(r)= v^2/ G r^2$. 
This leads to $\rho_\lam \approx \rho_E$ and  $GM/r\approx v^2$, so that $v(r)=v$ exhibits the flat rotation curve.  The acceleration $g$ consists of the Newton term $g_N=GM_b(r)/r^2$
of the baryons (stars and hydrogen clouds) enclosed within $r$, and the DM term; they combine essentially as  \citep{banik2022galactic}
\BEQ
g={\rm max}(g_N ,\sqrt{g_N \,a_0}\,\,),\qquad a_0=\frac{v^4}{GM_b} .
\EEQ
(The true MOND relation $g=g_N \, f(g_N/a_0),$  with $f(x)=\sqrt{x}$ at small $x$ and $f=1$ at large $x$, employed in applications is more rich than this form \citep{milgrom1983modification,lelli2017one,banik2022galactic}.)
It was supposed that $a_0\sim 1.2\,10^{-10}$m/s$^2$ is universal; empirical values fluctuate around this   \citep{rodrigues2018absence},
but confirm the baryonic Tully-Fisher relation $M_b\sim v^4$.

In our \EAE   approach, the amount of DM in a galaxy depends on its history, and there is no universal $a_0$.
The next subsection discusses that constant-density cores are favored in \EAEc, and next
we discuss various empirical evidence for that. 

Renzo's rule  states that ``for any feature in the luminosity profile'' (due to stars, X ray gas or hydrogen clouds)  
``there is a corresponding feature''  (a bump,  a dip) ``in the rotation curve and vice versa”  \citep{sancisi2004visible}.
The \EAE   approach offers a mechanism for this: extra structures in the electric field of extra mass structures.
However, data for the electric field is needed to give \EAE   predictive power.
Presently,  this is not available.

While MOND suffers from its conflict with wide binaries \citep{banik2024strong},  \EAEc  \,  is just Einsteinian/Newtonian  in this regime.

 \subsection{Stability analysis}
 \label{sec:stab}
 
 The stability of the theory needs to be considered. Following ref. \citep{nieuwenhuizen2023exact}, we
  consider a perturbation of the electrostatic potential and of the accumulated charge,
 \BEQ\label{delQrt}
 \delta A_0(r,t)=\eps a_0(r)e^{-i\om t}, \quad a_0'(r)=-j(r)E(r),\qquad
 \delta Q(r,t) =\eps \varrhoq(r) Q(r) e^{-i\om t},\qquad
 \EEQ
 with $0<\eps\ll1$ and yet unspecified profiles $\varrhoq(r)$, while we again neglect the matter density, but not the charge density, keeping $N(r)=1$.
This induces radial fluctuations of the metric, $\delta g_{\mu\mu}(t,r)=2\eps g_{\mu\mu}(r)h_\mu(r)e^{-i\om t}$ with $h_3=h_2$ for spherical symmetry.
The Coulomb energy density picks up metric fluctuations, 
\BEQ  \label{rhoEsource}
\delta\rho_E(r,t)=\eps \rho_E^\heen(r)e^{-i\om t},\qquad
\rhoA ^\heen 
=2\frac{\varrhoq (r)-h_0(r)-h_1(r)}{N^2(r)}  \rho_E(r)
\EEQ
The remaining Einstein equations correspond to perturbations in the coefficients of  (\ref {TMnrhomA}), 
\BEQ
T^\mu_\ednu=(\bar\rho_\lam+\delta \bar\rho_\lam)\delta^\mu_\ednu+(\rho_E+\delta\rho_E)\cC^\mu_\ednu+(\sigma_\vth+\delta\sigma_\vth)U^\mu U_\nu,
\quad \delta \bar\rho_\lam=\eps\rho_\lam^\heen(r)e^{-i\om t},\quad 
\EEQ
where $U^\mu=\delta^\mu_0/\sqrt{g_{00}}$ with $U_\nu=g_{0\nu}/\sqrt{g_{00}}$  is the time-like unit vector. 
A first effect is the appearance of elements $G^0_{\ed1}\sim G^1_{\ed0}\sim\om$, which must vanish since there is no energy current.
(While the radial electric field is enhanced by a radial charge current, viz. $\dot\vE=-\vJ_q$, this does not generate a magnetic field.)
This imposes 
\BEQ\label{h1r=}
h_1=r h_2'+h_2-h_2 \frac{r\Se  '}{2 \bar \Se  }  ,\qquad \qquad  (\om\neq 0),
\EEQ
a relation best employed only near the end of the analysis.
The remaining Einstein equations correspond to the first order perturbations in the coefficients of  (\ref {TMnrhomA}), 
\BEQ
\delta\rho_\lambda=\eps\rho_\lambda^\heen(r) e^{-i\om t},  \quad
 \delta\rhoA =\eps\rhoA ^\heen(r) e^{-i\om t} .
 \EEQ

Since the baryons make up a minor part of the energy,  we first omit their mass and pressure,  neglecting $\sig_\rmm$ and $\delta\sig_\rmm$,
 but keeping their net charge distribution.  This sets $\Sn(r)=1$. One can eliminate $h_0(r)$ in favor of a nucleus $\gn(r)$, 
\BEQ
\gn=h_0+h_1-h_2-rh_2' ,
\EEQ
after which the $h_i$ functions  follow as
\BEQ \label{h012g0}
h_0=\gn -\frac{ \bar \Se  \Se  '} { 2\om^2 } \gn '  ,\qquad
h_1= -\frac{   3\bar \Se \Se  ' }{2 \om^2}\gn ' -\frac{\bar \Se^2  }{ \om^2}   \gn '' , \qquad 
h_2= -\frac{ \bar \Se  ^2} { \om^2 r} \gn '   .
\EEQ
This finally leads to an equation for $\gn$ sourced by $j$,
\BEQ && \hspace{-12mm} \label{gEeqn0}
 \bar \Se  ^2 \gn ''  +2 \bar \Se   \Se  ' \gn '   +\frac{E'}{E}\bar \Se  ^2 \gn ' - \om^2 ( \gn  -\varrhoq) =0 .
\EEQ
The $E'$  term enters through $S_1'''$ from $G^\mu_\ednu$, while the other ones  come from $T^\mu_{\em\, \nu}$.

In our nonrelativistic application we have  $S,S'\approx0$, $\bar S\approx-1$, so the leading terms are
\BEQ \label{h012g}
h_0=\gn +\frac{   \Se '  \gn ' } { 2\om^2 } ,\qquad
h_1= -\frac{\gn''}{\om^2} +\frac{ 3 \Se  ' \gn ' +4 \Se   \gn ''  }{2 \om^2} , \qquad 
h_2= 
-\frac{  \gn '} { \om^2 r}  +\frac{ 2 \Se  \gn '} { \om^2 r}   .
\EEQ
and
\BEQ  \label{gEeqn} 
 \gn '' +\frac{E'}{E}\gn ' -  \om^2 ( \gn  -\varrhoq)=0. 
\EEQ
Eq. (\ref{gEeqn}) has a well behaved inhomogeneous solution, related to charge relocation coded by $\delta Q \sim  j(r)Q(r)$ from eq. 
(\ref{delQrt}).
$h$ behaves at large $r$ as $\gn(r)=\varrhoq(r)[1+\cO(1/\om^2r^2)]$.  To determine it in general,
let us denote the two real valued homogeneous solutions  of (\ref{gEeqn}) as $J_E(r)$ and $Y_E(r)$. Their Wronskian reads 
\BEQ\label{WE=}
W_E(r)\equiv J_E(r) Y_E'(r)-J_E'(r)Y_E(r)=\frac{E_\ast}{E(r)} ,
\EEQ
where $E_\ast>0$ follows by fixing the amplitudes and signs of $J_E$ and $Y_E$.
There is an inhomogenous solution of (\ref{gEeqn}) that is well behaved at large $r$, reading 
\BEQ \label{gEeqninh}
\gn_\inh(r)= -\om^2\int_r^\infty \d u\, \frac{J_E(r) Y_E(u) -Y_E(r) J_E(u)}{ W_E(u)}\, \varrhoq(u) .
\EEQ

\subsection{An instability leading to explosive core growth}
\label{sec:instab}

In our picture, dark matter stems from ZPE that flows in from infinity, and thereby creating a small
mismatch between the charged particles so as to set up an $E$ field.
Does the inflowing DE fill up profiles  $\rho_E\sim 1/r^2$ and hence $M_\tot\sim r$ and  $\Se (r)\sim r^0$ at large $r$, 
and if so, how does this happen? 

\begin{figure}
 \centerline{ \includegraphics[width=8cm]{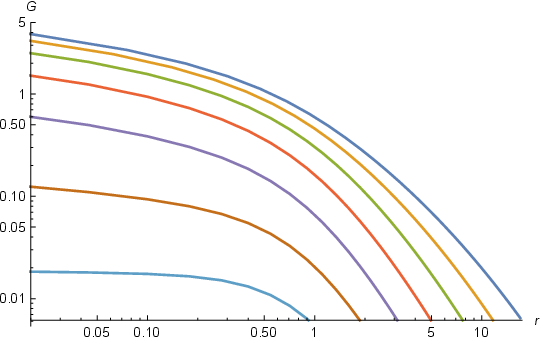}}
\caption{
The growth factor $G(r)$ of the mass  density in eq. (\ref{G(r)=}) for a toy galaxy with only $\nu=-1$ ($n=0$, $\rho_q={\rm cst}/r^2$)  dark matter
and enclosed-charge growth function $j(r)$ $=$ $1/\sqrt{r^2+1}$.
The temporal growth rates are $\varpi=2^k$ with $k$ from $-3$ (top) to $3$ (bottom).  }\label{figgp2r}
\end{figure}

To investigate this,  we assume $\Se (r) = 2v_n^2 r^{2n}$ for small and moderate $r$, for which  the rotation speed equals $v(r) =  v_nr^n$.
While  $\rho_\lam =(1+n) (2n+1) v_n^2 r^{2n-2}/8\pi G$, the related $\rho_E =(1-n) (2n+1)$ $v_n^2 r^{2n-2}/8\pi G$  must be
non-negative, allowing $-\half \le n \le 1$, with $\rho_E\to0$ for $n\to1$. 
For small $r$, our interest is $0\le n <1$ where $S\ll 1$ vanishes for $r\to0$ (it remains bounded for $n=0$).  
For relatively large $r$, the regime of interest is $-\half < n\le 0$.
On this background, we consider an electric field perturbation $\delta E(r,t) =\eps j(r) E(r) e^{-i\om t}$ with $\eps\ll1$ and some profile $j(r)$.
With the connection $\rho_E=E^2/8\pi= Q^2(r)/8\pi r^4$ this is a special case of the treatment in subsection \ref{sec:stab}, with $E\sim r^{n-1}$.
Eq. (\ref{gEeqn}) now reads
\BEQ
\gn''+\frac{1+2\nu}{r}\gn'+\om^2j=\om^2\gn,  \qquad 
 \nu = \frac{n}{2} -1 ,   \quad -\frac{5}{4}\le \nu\le -\half .
\EEQ
For real $\om$,  this has modes oscillating in time, involving modified Bessel functions $I_\nu(\om r)$ and $K_\nu(\om r)$.
More interesting are the unstable (growing) modes $j(r)e^{\varpi t}$, $\gn(r) e^{\varpi t}$ with $\varpi>0$ and $\om=i\varpi$, where
eq. (\ref{delQrt}) involves $\delta Q(r,t) =\eps \varrhoq(r) Q(r) e^{\varpi t}$.
The case $j(r)>0$ connects to  increase of positive charge in the  region $r<R_\co$,  
making the region $r>R_\co$ more negatively charged; the case $j(r)<0$ describes the reverse. Eq. (\ref{gEeqninh}) reduces to
\BEQ 
\gn_\inh(r)= \frac{\pi }{2}\varpi^2r^{-\nu}  \int_r^\infty \d u\, [J_\nu(\varpi r) Y_\nu(\varpi u) -Y_\nu(\varpi r) J_\nu(\varpi u) ] u^{\nu+1} j(u),
\EEQ
which involves the  ordinary Bessel functions $J_\nu$ and $Y_\nu$.
Compared to eqs. (\ref{WE=}) and (\ref{gEeqn}), it holds that $J_E(r)=r^{-\nu}J_\nu(\varpi r)$ and $Y_E(r)=r^{-\nu}Y_\nu(\varpi r)$, 
while $W_E(u)=2/\pi u^{1+2\nu}$.
The homogeneous solutions $J_E$ and $Y_E$ behave as $r^{(1-n)/2}\cos( \varpi r-\phi)$ at large $r$ and  have to be omitted.
Conversely, $\gn_\inh$ is well behaved and does not oscillate. 
With $\delta\rho_\lam/\rho_\lam=\delta\rho_E/\rho_E$, it follows that the increase of energy is exponential in time,
\BEQ\label{G(r)=}
\delta\rho_E \pplus \delta\rho_\lam = \eps (\rho_E \pplus \rho_\lam)
G(r)e^{\varpi t} ,\qquad 
G(r)
 = \frac{2 E' \gn_\inh'(r)}{\varpi^2E}
 = \frac{2 (n-1)\gn_\inh'(r)}{\varpi^2r}.
\EEQ
With $n\le1$ and,   if $j(r)>0$, $\gn'_\rmi<0$, it is seen that $G(r)>0$ describes mass accretion triggered by the increasing enclosed charge.
The profile $G(r)$ is plotted in Fig.  \ref{figgp2r} for the case $j(r)=1/\sqrt{r^2+1}$ and $n=0$, so that $\rho_q={\rm cst}/r^2$.
It shows that the ``explosive'' filling of the DM profile is most effective in the center.

It came as a surprise to us that MOND, corresponding to $n=0$ at large $r$, is not a limiting case, the allowed interval being $-\half \le n \le 1$.
Actually, the $n=0$ case at small $r$,  corresponding to a cusp $\rho_q\sim 1/r^2$, $E(r)\sim 1/r$,
$\rho_E\sim\rho_\lam\sim 1/r^2$,  is an extremal case \citep{nieuwenhuizen2024aether}.

There is evidence for the $n=1$, constant-density case, see next section.

When $j(r)<0$ in the galaxy or cluster center, the above argument exhibits a case of \EAE   flowing out of the core.
This may be connected with expanding and diluting cores of galaxies, 
see sec. \ref{reverse instab},  and  the cooling of cluster cores.

\subsection{Evidence for constant-density, non-cusped cores}
\label{sec:galcore}

The NFW profile for $\Lam$CDM \citep{navarro1997universal} has an $1/r$ singularity at the origin, 
expressing the mutual gravitational attraction of the CDM particles, playing out at their  low (``cold'') speeds.  
Whether the CDM consists of axions, WIMPs or MACHOs is not relevant  on the scale of galactic cores.
The $1/r$ divergence is called a cusp. However, dark matter cores are often observed to be flat, and the issue is called the ``cusp-core problem";
for recent reviews, see \citep{del2021review,boldrini2021cusp}. 
Dark matter halos seem to have stopped growing:
those of nearby quasars  are not heavies than those at $z\sim6$ \citep{arita2023subaru}. 

For \EAEc,  eq. (\ref{rhodm=}) typically involves  a constant-density DM core.

In galaxies, the nearly flat rotation curves occur since normal matter (stars, gas clouds) at small $r$ 
adds to the DM which dominates at large $r$. 
In NGC 3626  the rotation curve keeps growing beyond 8 kpc, for data up to 18 kpc.
After accounting for stars and clouds, ref.   \citep{shelest2020spirals}
adds a DM component with $v_\dm\sim r^{n}$ for $n\approx 1$ or $n = 1$ in 3 galactic models. 
The same is done for NGC 2824 and NGC 6176, which do show flattening. 
The DM with index close to or equal to our limit $n_\max=1$ supports the above analysis.

In the Triangulum galaxy (M33, Cartwheel galaxy) the rotation speed  increases  as constant+linear beyond 3 kpc up to 15 kpc. 
It was modelled by determining the contribution from stars and gas, while DM was modeled by an NFW profile \citep{lopez2017radial}.
As in previous cases, modelling by an $n=1$ (constant density) \EAE   profile may work.

A variety of observations are at odds with the presence of a cusp \citep{palunas2000maximum,salucci2000dark,deblok2001mass,karukes2017universal,di2019universal};
they favor a constant-density core of a few kpc in size. 
Ref. \citep{di2019universal} mentions a mysterious entanglement between the properties of the luminous and the dark matter, 
which has similarity to Renzo's rule of section \ref{sec5.1}. In \EAE   theory this occurs, since its ZPE dark matter is regulated by
an electric field, due to a net charge distribution that has to adjust itself.

Our \EAE   explanation for the DM in the Galactic center is an indirect support for the binary milisecond pulsar interpretation 
of the 511 keV Fermi-LAT line \citep{bartels2018galactic}.

\subsection{Dissolution of galactic cores}
\label{reverse instab}

Another piece of evidence is the observational evidence of evolving constant-density dark matter profiles \citep{sharma2022observational}.
By subtracting the contributions from normal matter, these authors study the dark matter halos of a set of 256 star-forming disk-like galaxies 
at redshift $z\sim 1$. They find constant-density DM cores, as expressed by eq. (\ref{rhodm=}) and corresponding to our above case $n=1$.
But, statistically, the DM cores at $z\sim1$ are denser by 1.5 dex  than current ones and smaller by a factor of 0.3 dex compared to present.

Within \EAE   theory, this can be explained by a diminishing of the net charge ratio, 
related to the instability of cores mentioned in sec. \ref{metastab-instab}.
It leads to  ZPE moving outwards after having moved inwards;
the same effect may also have caused the adiabatic expansion of the cluster gas in cool cores in clusters.
In principle,  this outflow can be described by an approach like the one in sec. \ref{sec:instab}, with negative source function $j(r)$, 
underlining once more the ZPE's fluid character.

\subsection{The electric field scaffold and the vast polar structure}
\label{VPS}

It has been established that the Milky Way galaxy is surrounded by a vast polar structure of subsystems: 
satellite galaxies, globular clusters and streams of stars and gas, spreading from Galactocentric distances as small as 10 kpc out to 250 kpc
 \citep{metz2007dwarf,pawlowski2012vpos,pawlowski2014co}. A similar structure  occurs in Andromeda  \citep{kroupa2014planar}.
While an explanation was given as tidal tails of material expelled from interacting galaxies, the predicted \EAE   scaffold offers a fresh viewpoint.
It presents a rather strong structure with which the matter inside it correlates.
This picture offers an alternative for the fortunate temporal alignment
 put forward recently for Gaia data interpreted within $\Lam$CDM \citep{sawala2023milky}.

\renewcommand{\thesection}{\arabic{section}}
\section{Electro-aether-energy  in clusters}  
\setcounter{equation}{0}
\renewcommand{\theequation}{7.\arabic{equation}}

\label{EVE-clusters}

\subsection{Modified isothermal sphere as a  fit for lensing}
\label{IsoLensingfit}

To apply the idea of isothermal spheres to the galaxy clusters, we consider the clusters A1689 and A1835. 
In ref.  \citep{nieuwenhuizen2021accurate}, precise strong lensing and gas data and fits to them were presented.
Here we consider a different modelling for the DM.
A regularization of an isothermal sphere is
\BEQ\label{depletedisothermal}
\rho_E=\frac{ E^2(r) }{8\pi} 
=\frac{v^2}{8\pi G}\frac{r^2}{r^2+r_0^2} \frac{1}{r^2+r_1^2} \frac{r_2^{2n_\co}}{(r^2+r_2^2)^{n_\co}} ,
\EEQ
with $r_0\ll r_1\ll r_2$.  At small $r$ it yields $E(r)\sim r$ and a finite central charge density $\rho_q(0)$. 
In the middle region, $r_0\ll r\ll r_2$, it acts as a truncated isothermal sphere. 
At large $r\gg r_2$, it exhibits incomplete build up (underfill) with index $n_\co$.

The data for the cylindrical mass $M_{2d}(r)$ are expressed in the cylindrical mass density $\bar \Sigma(r)=M_{2d}(r)/\pi r^2$,
which derives from the $3d$ mass density as
\BEQ \label{Sb1int2}
\bar\Sigma(r)=\frac{4}{r^2}\int_0^r\d s\,s^2\rho_\tot(s)+\int_r^\infty\d s\,\frac{4s\rho_\tot(s)}{s+\sqrt{s^2-r^2}} . 
\EEQ

\begin{figure}
\centerline{ \includegraphics[width=8cm]{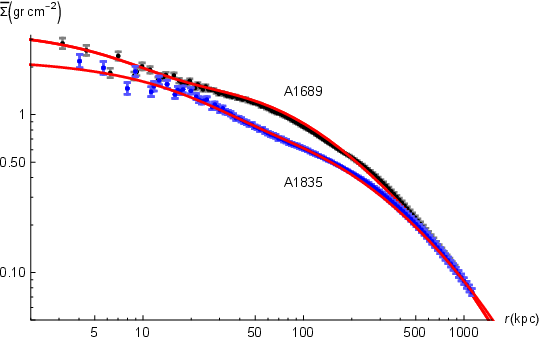}}
\caption{
Cylindrical mass density as function of $r$ in the galaxy clusters A1689 (upper) and A1835 (lower)
with their fits to the truncated isothermal profile of Eq. (\ref{depletedisothermal}) with index $n_\co=2$.
}\label{figSiBDM=DE}
\end{figure}

\noindent 
We consider the profile (\ref{depletedisothermal}); the case $n_\co=2$ works well;  an analytic expression for its contribution to $\bar\Sigma$
can be derived. For the brightest cluster galaxy (bcg) with mass $M_\bcg$, we add a stretched exponential (se) profile 
$\rho_{\rm bcg}=[M_{\rm bcg} /4\pi \Gam(3n_\se) n_\se R_\se^3] $ $\times$ $\exp[-(r/R_\se)^{1/n_\se}]$,
while the X-ray gas has also been modelled in  \citep{nieuwenhuizen2021accurate}.
In Fig.  1 we present the data for $\bar\Sigma(r)$ and fit this to $\bar\Sigma_\bcg+\bar\Sigma_\dm+\bar\Sigma_{\rm gas}$
with $r_0\to 0$.
The further parameters are for A1689:
\BEQ
&& 
\hspace{-12mm}
 M_\bcg= 1\,10^{12}M_\odot ,\hspace{1mm}
R_\se = 3 \, \kpc, \hspace{1mm}
n_\se = 1 , \hspace{1mm} 
v=3480 \frac{\km}{\rms} , \hspace{1mm}
 r_1 = 50\,\kpc,\hspace{1mm}
r_2 = 1.5 \, \Mpc ,\hspace{1mm}
\EEQ
and for A1835:
\BEQ
&& \hspace{-14mm}
 M_\bcg = 9\,10^{12}M_\odot ,\hspace{0.5mm}
R_\se  = 6 \, \kpc, \hspace{0.5mm}
n_\se = 1.25 , \hspace{0.5mm}
 v = 3350 \frac{\km}{\rms} ,  \hspace{0.5mm}
 r_1 = 100\,\kpc,\hspace{1mm}
r_2 = 2.1 \, \Mpc .
\hspace{1mm} 
\EEQ
These values work well, but are not optimized;  the error bars will be comparable to the ones  in related fits 
 \citep{nieuwenhuizen2021accurate}. Since the bcg is poorly constrained, other shapes may function as well.
The ``underfill''  for $r\gtrsim  r_2$ expresses that the surplus electrons pushed outwards
are dominant  there and diminish the net enclosed charge $Q(r)$ and hence $\rho_E$. 

According to  (\ref{rhol=}), the LCC $\rho_\lam(r)$ equals $\rho_E^>(r)-\rho_E(r)$. For $\rho_E$ in eq. (\ref{depletedisothermal}) with $n_\co=2$, 
the $\rho_E^>$ term can be solved analytically, 
\BEQ \hspace{-3mm}
\rho_E^>(r)=4\int_r^\infty \d u\frac{\rho_E(u)}{u}=\frac{r_2^4 v^2 }{4 \pi  G}
\left(
\frac{1}{r_{20}^2 r_{21}^2 \left(r^2+r_2^2\right)}
-\frac{L_0}{r_{10}^2 r_{20}^4}
+\frac{L_1}{r_{10}^2 r_{21}^4}
-\frac{r_{20}^2+r_{21}^2} {r_{20}^4 r_{21}^4} L_2\right),
\EEQ
with $r_{10}^2=r_1^2-r_0^2$, $r_{20}^2=r_2^2-r_0^2$, $r_{21}^2=r_2^2-r_1^2$ and $L_i=\log(r^2+r_i^2)$.
As in the toy galaxy of sec. \ref{sec:toygal},  $\rho_\lam$ is positive for small and moderate $r$, while it has a negative tail $-r_2^4v^2/24\pi Gr^6$.
Its zero crossing lies at $1.2$ Mpc for A1689 and at $1.7$ Mpc for A1835.

The  crossover radius which separates the inner region with positive net charge density $\rho_q$ from the outer region with a negative one,
occurs for $\d (r^4\rho_E)/\d r=0$, implying $R_\co\approx r_2$ for $n_\co=2$,
so that $R_\co=1.5$ and 2.1 Mpc for A1689 and A1835,  respectively.

\subsection{The hydrostatic equilibrium puzzle in clusters}
\label{HEpuzzle}

In the Earth's atmosphere, hydrostatic equilibrium  is broken by lightning, after which its restoration leads to thunder. 
In studies of clusters,  hydrostatic equilibrium of the X-ray gas is investigated, but found to be dissatisfied
 \citep{morandi2010unveiling,lemze2011quantifying,morandi2012x}; it
 leads to a $\sim 40\%$ ``nonthermal pressure'' component in the center of A1689  \citep{molnar2010testing},
 supposedly due to turbulence, spurious gas dynamics or the dynamical build up of the cluster..
 Figure 5 of  our ref. \citep{nieuwenhuizen2013observations} shows that for hydrostatic equilibrium
  the gas temperature (and with it, the pressure) should be larger than observed by a factor $\sim 1.5$. 

This riddle can be solved in \EAE   theory.
The condition for hydrostatic equilibrium, Eq. (\ref{new}), reads
$p_\vth'+p_\lam'= \cFE +\cFG $, with
\BEQ \label{new3}
 \cFE =\rho_E'+4\frac{\rho_E}{r}, 
  \qquad 
\cFG \approx -G (\rho_\rmm+p_\rmm)\frac{M_\tot+4 \pi  r^3(p_\vth-\rho_\lam-\rho_E)}{r^2} .
\EEQ

\subsubsection{Hydrostatic equilibrium in $\Lam$CDM}

In $\Lam$CDM one has $\rho_E=0$ and $p_\lam(r)=-\Lam/8\pi G$, a constant;  employing 
$\rho_\vth=\rho_\bcg+\rho_g$ and 
$p_\vth=p_g$ from (\ref{rhompm=}),
eq. (\ref{new3}) results in 
\BEQ \label{he-lcdm}
p_g'\approx -G (\rho_\bcg+\rho_g)\frac{M_\tot}{r^2} , \qquad  \qquad \qquad  (\Lam\text{CDM}).
\EEQ
This  exhibits the $\Lam$CDM hydrodynamic equilibrium condition  for the gas pressure, but notice that $\rho_\vth=\rho_\bcg+\rho_g$
 also involves the bcg,  a point generally overlooked. 

A consistent approach in $\Lam$CDM simulations should satisfy (\ref{he-lcdm}), but observed clusters need not, and, as mentioned above,
the analyzed ones do not satisfy it. Away from the bcg, where $\rho_\bcg\to0$, the relation remains violated.

\subsubsection{Hydrostatic equilibrium in \EAE  }
\label{HEcluster-EVE}

In the \EAE   approach, the $\rho_E$ and $p_\lam$ terms are present with $|p_\lam|\sim \rho_E$, and they are by far the dominant terms,
 making (\ref{new3}) a relation between them with the matter terms as  spectators.
  As we realized in our fitting of the lensing profiles of the clusters A1689 and A1835
 \citep{nieuwenhuizen2013observations,nieuwenhuizen2017zwicky,nieuwenhuizen2020subjecting,nieuwenhuizen2021accurate}, 
the gas is just a spectator,  with right nor need for ``its own'' hydrostatic equilibrium.

 In the description of section \ref{linEineq} the nonlinear $\cFG $ terms were left out, which led to the
$\rho_\lam$ in eq. (\ref{roLa=}) as solution for hydrostatic equilibrium.  
$\cFG $ indeed acts as a small correction; compared to $\cFE $, it is of relative size $ 10^{-6}- 10^{-5}$.
Integrating it from $r$ to $\infty$ yields a correction to $\rho_\lam$, which
exhibits the malleability of the ZPE, doing just the right thing in the situation at hand.

The respect for hydrostatic equilibrium in \EAE   implies that no big effects of turbulence or other  (gas) dynamics are to be sought for.
Rather, it underlines that its violation in $\Lam$CDM is a real deficit.

\begin{figure}
\centerline{ \includegraphics[width=8cm]{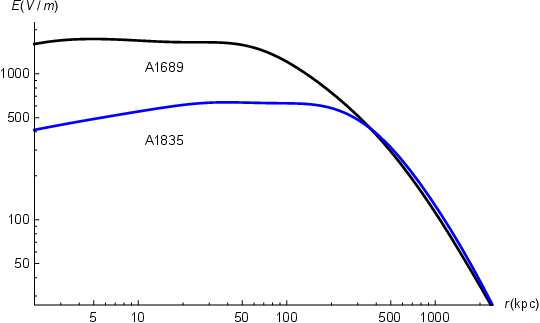}}
\caption{Prediction for the static electric field (in V/m) as function of the radius in the galaxy clusters A1689 (upper, black) and A1835 (lower, blue).
}\label{figstaticE} 
\end{figure}

While $\rho_\lam+\rho_E-p_\vth$ can be identified with the empirical $\rho_\dm$ profile,  
data for $\rho_E$ are needed to test the hydrostatic equilibrium condition. Unlike the ZPE density, that can only be inferred gravitationally,
 the electric field acts on charges, and can in principle be determined. 
Eq. (\refl{roLa=}) yields the prediction $E(r)$ $=[-r\rho{}'_\tot(r)/2\eps_0  ]^{1/2}$; it is plotted in fig.  \ref{figstaticE} for  A1689 and A1835.
The central regions involve $\sim1.5$ and $\sim0.5$ kV/m, respectively.

\renewcommand{\thesection}{\arabic{section}}
\section{Electro-aether-energy in cosmology}
\setcounter{equation}{0}
\renewcommand{\theequation}{8.\arabic{equation}}

\label{EVE-cosmology}

\subsection{Zero pressure \EAE   equation of state}
\label{p=0-eqn-of-state}

On cosmological scales, one considers the spatial average of the mass density and pressure over a large cosmological volume $V$ with many galaxies.

For a given galaxy (cluster),   the electrons compensate the positively charged interior of the galaxy (cluster) 
beyond the crossover length $R_\co$, making  $\rho_E$ decay faster than  $1/r^4$. 
The integral in eq. (\ref{Mtotr=}) vanishes in the limit $r\to\infty$ , so as to keep the limits
\BEQ
M_\lam+M_E+M_\vth^\rho=\frac{4}{3}M_E +M_\vth^p .
\EEQ
Likewise, consider the energy stored in the pressures,
\BEQ
P_\tot^i=4\pi \int_0^\infty\d u\, u^2p_\tot^i(u),\quad i={r,\theta,\phi} .
\EEQ
Eq. (\ref{pitot}) yields the values
\BEQ \label{Pvals}
P_\tot^{r}=-\frac{4}{3}M_E,\quad P_\tot^\theta=P_\tot^\phi=P_\tot^\perp=\frac{2}{3}M_E.
\EEQ

A large cosmological volume $V$ with many galaxies at random positions has a mass density and an isotropic pressure. 
The mass density can be obtained by smearing out the mass of each galaxy over V and summing.
The $3\times3$ pressure matrix is, on the average, diagonal and isotropic.
For each galaxy,  the contribution to the pressure is obtained as $1/3V$ times the trace of the pressure matrix integrated over space.
Eq. (\ref{Pvals}) implies that this trace vanishes for each galaxy and hence for their combination. 
This leads to the equation of state for the \EAE  
\BEQ\label{pezp=0}
p_\ezp=0 , \qquad w_e \equiv \frac{p_e}{\rho_e}=0,
\EEQ
which coincides with the standard $p_\dm=0$ for particle cold dark matter like $\Lam$CDM.

In an alternative approach, we only use that the trace of the EM stress energy tensor vanishes. Averaging over a large volume $V$
with many galaxies yields an isotropic pressure $p_E=\frac{1}{3}\rho_E$. With $p_\zpe=-\rho_\zpe$ and $\rho_\zpe=\frac{1}{3}\rho_E$\footnote{A factor 
3 between electric and zero point contributions was  first encountered in integral form in the exact solutions
 for black holes with a regular interior \citep{nieuwenhuizen2023exact}, and also occurred in the present work at the end of sec. \refl{sec:toygal}.}
, eq.  (\ref{pezp=0}) follows again from $p_e=p_\lam+p_E$.

\subsection{The Hubble tension and the increasing amount of dark matter}

The so-called Hubble tension is the fact that local measurements of the Hubble constant via supernovas yield 
$H_0\approx 73$ km/s Mpc \citep{brout2022pantheon}, while the cosmic microwave background fixes it at $\approx68$ km/s Mpc \citep{aghanim2020planck}, with smaller and smaller error bars bringing it to a $\sim5\sigma$ discrepancy. 
It is often argued that late time physics cannot solve the Hubble tension, see, e.g.,  
the ``most general'' scenario \citep{keeley2023ruling}.
The reason for this is simple, at the present low temperature of the cosmos, no new dark matter particles can be created,
so there cannot be an increase of cold dark matter. 
But it is also early-time new physics alone cannot solve the Hubble tension \citep{vagnozzi2023seven}. 
Apparently, the situation is presently  in a limbo.

While the \EAE   acts as a  pressureless type of cold dark matter,
its {\it mass  density can grow in time}, since more and more ZPE/AE  can be condense {\it locally} on BHs, galaxies, clusters, filaments, etc.
This will lead to a growth equation for the cosmic dark matter fraction $\Om_\dm$; a modelling is given in eq. (\ref{Omdme}) below.
It fixes the ``enslaved'' {\it global} dark energy density $\Om_\lam$ by energy conservation, see eq. (\ref{Omlama}) below.

\subsection{{ Friedmannology for aether energy condensation}}
\label{Hubble-tension}

\newcommand{\Oma}{\Om}
\renewcommand{\Oma}{E}

The Friedman equations for the cosmic scale factor $a(t)$ with $a({\rm now})=1$, are
\BEQ \label{Friedman}
H^2=\frac{\dot a^2}{a^2}=\frac{8\pi G}{3}\rho,\quad \frac{\ddot a}{a}=-\frac{4\pi G}{3}(\rho+3p),\quad \d(\rho a^3)=-p\d a^3.
\EEQ
Two of these three equations are independent; the last one expresses energy conservation.
It is customary to divide out the critical density $\rho_c=3H_0^2/8\pi G$, where the Hubble constant $H_0$ is the present value of  $H,$
and to split up in various components $\Om_i=\rho_i/\rho_c$,
\BEQ\label{Friedmon1}
\frac{H^2}{H_0^2}=\Oma(a),\qquad 
\Oma(a)=\sum_i\Om_i(a).
\EEQ
We will consider radiation ($r$), baryons ($b$), dark matter ($dm$), and aether energy ($\lam$). 
\BEQ 
\Oma(a)=\Oma_r(a)+\Oma_b(a)+\Oma_\dm(a)+\Oma_\lam(a)\equiv \frac{\Om_r(a)}{a^4}+\frac{\Om_b(a)}{a^3}+\frac{\Om_\dm (a)}{a^{3}}+\Om_\lam(a) .
\EEQ
With $w_\dm\equiv w_e=0$, see eq. (\ref {pezp=0}),  the respective equation of state parameters are
\BEQ  
w_r=\frac{1}{3},\qad w_b=w_\dm=w_e=0,\qad w_\lam=-1 .
\EEQ
Energy conservation gets expressed as 
\BEQ
\label{Friedmon}
\Om_\lam'=-\sum_{i\neq\lam}\Big[ \Oma_i'(a)+3(1+w_i)\frac{\Oma_i(a)}{a} \Big]=-\frac{\Om_r'}{a^4}-\frac{\Om_b'}{a^3}
-\frac{\Om_\dm'}{a^3} , \quad
\EEQ
with the solution
\BEQ \label{Omla,Ea}
&& \hspace{-10mm}
\Om_\lam(a)=\Om_\Lam+\int_a^1\d b\Big[\frac{\Om_r'(b)}{b^4}+\frac{\Om_m'(b)}{b^3}+\frac{\Om'_\dm(b)}{b^3}\Big] ,  \\ &&
\hspace{-10mm}
E(a)=1+\int_a^1\frac{\d b}{b}\Big[4\frac{\Om_r(b)}{b^4}+3\frac{\Om_b(b)}{b^3}+3\frac{\Om_\dm(b)}{b^3}\Big],\qad 
\quad \Om_\Lam+\Om_r(1) +\Om_b(1) +\Om_\dm(1)=1. \nn
\EEQ
In the present epoch, $\Om_r(a)=\Om_r$ and $\Om_b(a)=\Om_b$ are constants, but they change when neutrinos become nonrelativistic
due to their small, finite masses; in the past such changes happened during the freeze out of the various species, accompanied by a change in $\Om_\lam$.

Deviating from the standard assumption that  also $\Om_\dm$ and $\Om_\lam$ are constants,  we will consider an increasing $\Om_\dm(a)$ 
with an appropriate, non-constant $\Om_\lam(a)$.

The prediction of a growing $\Om_\dm$ is supported by observational values for the cosmic matter fraction $\Om_\rmm=\Om_\dm+\Om_\rmb$.
The early time value $0.315\pm 0.007$ from the cosmic microwave background (CMB)
observed by the Planck satellite \citep{aghanim2020planck},  is smaller than the late time (``now'') value $0.334 \pm0.018$ deduced  
from supernovae in the nearby cosmos  \citep{brout2022pantheon}. 

To deal properly with the problem, a unified approach covering these epochs is needed. 
Here we connect to ref.  \citep{dainotti2021hubble}. By parting the Pantheon supernova data in redshift bins, 
a weak time-dependence of Hubble constant $H_0$ is  found and fit to the form
$H_0(z)=H_0^\now(1+z)^{-\alpha}$ with $\alp\approx 0.01$.
Inspired by this,  we consider the DM growth function  
\BEQ \label{Omdme}
\Om_\dm(a) =\Om_ea^{\dele} ,
\EEQ
 where $\dele\sim 0.02$, deviating from $\dele =0$ in $\Lam$CDM. (For simplicity, we neglect a possible $a$-dependence of $\dele$).
This small value is compatible with the non-growth of  dark matter halos of quasars since $z\sim6$ \citep{arita2023subaru}. 

So, next to radiation ($r$), baryons ($b$), we consider \EAE   ($e$)  dark matter, and a time dependent dark energy ($\lam$). 
The  respective equation of state parameters are
\BEQ  
w_r=\frac{1}{3},\qad w_b=w_e=0,\qad w_\lam=-1 .
\EEQ
Since $\Om_b$ and $\Om_\ezp$ thus have negligible pressure, the total pressure reads 
\BEQ
\frac{p}{\rho_c}=\frac{\Om_r}{3a^4}-\Om_\lam(a) .
\EEQ
The total energy content is now
\BEQ \label{Omtot}\label{Etot}
\Oma(a)=\frac{\Om_r}{a^4}+\frac{\Om_b}{a^3}+\frac{\Om_\dm (a)}{a^{3}}+\Om_\lam(a)
=\frac{\Om_r}{a^4}+\frac{\Om_b}{ a^3}+\frac{\Om_e a^\delta_e}{a^3}+\Om_\lam(a) . \quad 
\EEQ
Like  in other applications, the aether,  here expressed in the component $\Om_\lam$,  is enslaved.
Eq.(\ref{Friedmon}), expressing energy conservation,  fixes it as
\BEQ \label{Omlama}
\Om_\lam(a)=\Om_\Lam+\frac{\del_e}{3-\del_e}\Om_e (a^{\del_e-3}-1),\qad
\EEQ
with $\Om_\Lam$ the present cosmological constant. The $a$-depencence exhibits  that $\Om_\lam(a)$ was much larger in the past than now,
and will be smaller in the future: Throughout the history of the Universe, aether energy is turned into other forms of energy.
In this setup, the smallness of $\Om_\lam(1)=\Om_\Lam$ is not a result of fine-tuning, but of dynamics, set by the {\it integration constant} $\Om_\Lam$.

Eqs. (\ref{Etot}) and (\ref{Omlama}) lead to
\BEQ \label{Oma3=}
\Oma(a)
=\frac{\Om_r}{a^4}+\frac{\Om_b }{a^3}+\frac{3a^{\delta_e-3}-\dele}{3-\dele}\Om_e +\Om_\Lam ,
\EEQ
which can be written in the effective form
\BEQ \label{Omafin}
&& 
\Oma(a) =\frac{\Om_r}{a^4}+\frac{\Om_b }{a^3}+\frac{\bar\Om_e}{a^{3-\delta_e}}+\bar\Om_\Lam
=\frac{\Om_r}{a^4}+\frac{\Om_b }{a^3}+\frac{\bar\Om_\dm(a)}{a^{3}}+\bar\Om_\Lam,
\\&& 
 \bar\Om_\dm(a)=\bar\Om_e a^{\dele} ,  \qquad 
\bar\Om_e=\frac{3\Om_e}{3-\dele} , \qquad \bar\Om_\Lam=\Om_\Lam-\frac{\dele \Om_e}{3-\dele} , \nn
\EEQ
with, as usual, the present-time sum rule
\BEQ
\Oma(1)=\Om_r+\Om_b+\Om_e+\Om_\Lam=\Om_r+\Om_b+\bar\Om_e+\bar\Om_\Lam=1 .
\EEQ
Eq. (\ref{Omafin}) has the familiar form, only modified by an increasing dark matter component.
Unlike other rather ad hoc approaches, our ongoing \EAE   condensation  Ansatz (\ref{Omdme}) fits in a bigger picture, 
the one where \EAE   also acts as the dark matter in galaxies and clusters, as ingredient of singularity-free black holes, and more.

\subsection{A fit to CMB data}

Identifying $H_0^\cmb\equiv z_\cmb^{-3/2}[H(z_\cmb)]^{1/2}$ leads to 
 \BEQ
 H_0^\cmb \approx H_0^\now\Big( \frac{\Om_b+\Om_\dm z_\cmb^{-\dele } }{\Om_b+\Om_\dm} \Big)^{1/2} ,
 \EEQ
  with $z_\cmb\approx 1080$.
 The value $\del_e= 0.025$  maps $H_0^\now=73$ km/s Mpc to $H_0^\cmb\approx 68 $ km/s Mpc, apparently solving the  Hubble tension
with late-time physics, an option unjustly ruled out by restricting  in ref.  \citep{keeley2023ruling} to the ``most general'' scenario.

Of course, a more fundamental analysis is warranted. Employing the CLASS code  \citep{blas2011cosmic,di2013classgal}, it is possible to find reasonable
parameter fits to the Planck CMB TT, EE and TE spectra \citep{aghanim2020planck} up to angular index $l=2000$.
A specific case
\footnote{$H_0=74$, $\dele =0.015$, $\tau=0.1225$, $\om_b= 0.0231$,  $\om_c= 0.128$, $\ln 10^{10}A_s= 3.162$, $n_s= 1.029$, 
$Y_{\rm He} = 0.2398$, $N_{ur}= 3.719$, $\Om_k =0.00466$,  $\chi^2/\nu=1.177$.
 \label{myCMBfit}}
 is depicted in fig. 1.   While encouraging by its good fit, this is only indicative.
 In a proper approach one has to derive the theoretical CMB spectra for the \EAE   situation of (\ref{Omdme}),
considering effects of the electric fields, 
and fit those predictions to the various data sets, such as CMB, baryon acoustic oscillations  and supernovae. 
At the next level of description, one determines a practical shape for $\Om_\dm(z)$ from Monte Carlo simulation or otherwise.
These steps are beyond the aim of the present paper;  we restrict  ourselves to stating that the Hubble tension is eased
and likely solvable in \EAEd.

\subsection{The large Hubble constant and the age of the Universe}

Though too seldomly stressed, a large Hubble constant leads to a small age of the Universe, bringing it close to physical lower bounds.
The value $H_0=73$ km/s Mpc leads to
\BEQ
\text{Age of Universe} \approx  12.8 \text{  Gyr},
\EEQ
 rather than the 13.8 Gyr in $\Lam$CDM.
This is slightly older than the age of the oldest cluster M4
based on main sequence stars $12.6\pm 1.1$ Gyr, or based on the oldest  white dwarfs $12.7 \pm 0.7$ 
Gyr \citep{rich2009fundamentals}.
But it is younger than the estimated age  $13.7\pm 0.7$ Gyr for the Methusalah star HD 140283  \citep{creevey2015benchmark},
and the accurate $13.535 \pm 0.002$ Gyr of the ultra-metal poor 2MASS J18082002–5104378 B \citep{schlaufman2018ultra}.

These and related estimates are based on fitting to   simulations of  $\Lam$CDM, not to \EAE   theory.
Its predicted early structure formation is suited for early objects. 
Anyhow, in no consistent theory an age larger than its age of the Universe should appear.

\begin{figure} 
 \centerline{ \includegraphics[width=5.3cm]{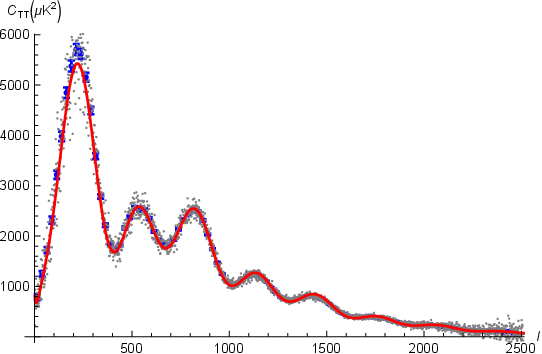} \includegraphics[width=5.3cm]{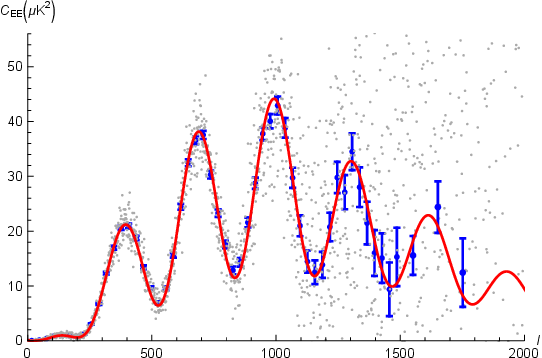} \includegraphics[width=5.3cm]{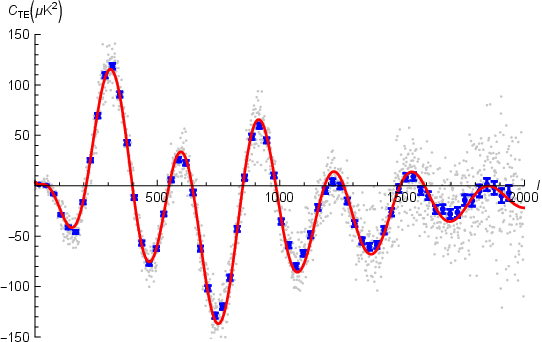}}
\caption{Fits to the TT, EE and TE spectra of the cosmic microwave  background observed by the Planck satellite for $l$ up to 2000.
The large value of the Hubble constant of 73 km/s Mpc is compatible with the data due to the dynamical nature of electro-zero-point energy.
 }\label{figBr}
\end{figure}

\subsection{The lopsidedness of the cosmos and the axis of evil}

 Another challenge to the standard cosmological model  concerns the cosmological principle: the expansion of the Universe is homogeneous and isotropic. 
 The largest effect in the cosmic microwave background fluctuations is the so-called the dipole asymmetry.
 Is it simply due to the motion of the Galaxy through the cosmos, or is it due to a genuine asymmetry in the distribution of matter? 
 Analysis of X ray galaxy clusters \citep{migkas2020probing}  and radio galaxies and quasars \citep{secrest2022challenge}
 suggests that the universe is lopsided in our frame. This correlates with the mysterious ``Axis of evil'' \citep{land2005examination}, the fact that
the plane of the  Galaxy correlates with the alignment of the low-$l$ ($l=2,3,4,5$) multipoles of the CMB: the ``top half'' of the cosmos is sllghtly 
cooler than the ``bottom half'', and the axes of the quadrupole and octopole correlate with it.
 
 It is natural to imagine that the cosmic expansion occurred  in an anisotropic way, and, consequently, 
 that the cosmological constant was anisotropic. \EAE theory can accommodate that, since  it connects the  cosmological constant
 (as vacuum energy) to matter,  via the necessary electric fields carried by charge mismatches.

\subsection{Beyond present}

Except for special periods in the early Universe, $\Om_r'$ and $\Om_b'$ are zero,
so that $\Om_{r,b}(a)=\Om_{r,b}$ keep their present values in the future. (For simplicity, we neglect
the fact that neutrinos have a small mass and that BHs radiate).
 If $\Ome (a)$ were also constant, we would have the $\Lam$CDM connection $\Om_\lam(a)=\Om_\Lam$.
Since $\Ome(a) =a^3E_\dm(a)$ is the cosmic \EAE   fraction in a comoving volume, it can continue to increase by further condensation as DM.

Let us consider the far future where expansion leads to a scale factor $a>1$ or even $\gg1$ 
and the integral in the expression for $E(a)$ in (\ref{Omla,Ea}) is negative.
The \EAE   condensation will likely go on until the dissolution of \EAE   in galaxies and clusters possibly takes the overhand, which could be modeled 
by a parameter $\del_\ezp<0$. 
Finally this leads to a ``true'' cosmological constant,
\BEQ
\Om_\Lam^\ezp\equiv \Om_\lam(a_{\rm max})=\Om_\Lam-\int_1^{a_{\rm max}}\d b \frac{\Ome '(b)}{b^3}.
\EEQ
with finite or infinite $a_{\rm max}$.
In the latter case, it results in 
\BEQ\label{Oml,Ea}
 \Om_\lam(a)=\Om_\Lam^\ezp+\int_a^\infty \d b \frac{\Ome '(b)}{b^3} ,\quad
 \Oma(a)=\Om_\Lam^\ezp +\frac{\Om_r}{a^4}+\frac{\Om_b}{a^3}
 + 3 \int_a^\infty\frac{ \d b}{b} \frac{\Ome (b)}{b^3}.
\EEQ
The proposed shape $ \Ome (a)=\Ome  a^{\dele }$ leads for $\dele <3$ to
\BEQ
\Om_\lam(a)=\Om_\Lam^\ezp+\frac{\dele \, \Ome }{3-\dele }\frac{a^{\dele }}{a^3} , \quad 
\Oma(a)=\frac{\Om_r}{a^4}+\frac{\Om_b}{a^3} +\frac{3 \Ome }{3-\dele }\frac{a^{\dele }}{a^3}+\Om_\Lam^\ezp .
\EEQ
The case $\dele =1$ connects to $\Ome (a)/a^3=\Ome(1) /a^2$ which is commonly connected to curvature of space; 
here it is a special -- and relatively large --  parameter $\dele$.

The deceleration parameter defined as
\BEQ\label{q(a)=}
q=-\frac{a\ddot a}{\dot a^2}=-1-\frac{aE'(a)}{2E(a)} .
\EEQ
The present value, 
\BEQ\label{q0=}
q_0=-1+2\Om_r+\frac{3}{2}\Om_b+\frac{3}{2}\Ome\approx -0.55,
\EEQ 
exhibits acceleration ($q_0<0$). It coincides with the Planck value \citep{aghanim2020planck}.
Analysis of the Pantheon supernovae sample \citep{camarena2020local,camarena2020new} leads, however, 
to $q_0=-1.1\pm0.3$, disagreeing at $2\sigma$. The Dark Energy Survey finds 
$q_0 =-0.530^{+0.018}_{-0.017}$ \citep{abbott2024dark}, at 1.5$\sigma$ from (\ref{q0=}).

\subsubsection{Black holes and the big crunch}

Ref.  \citep{farrah2023observational} considers that supermassive BHs 
in the redshift interval $0.7<z<2.5$ have a mass $M=M_i(a/a_i)^k$ growing with the scale factor as $a^k$ since 
time $t_i$ where $a(t_i)=a_i$, where $a=1/(1+z)$,   and construct a histogram of the distribution $p(k)$, normalized as $\int\d k\,p(k)=1$. 
Being centred around $k = 3$, it appears to have weight between $k_\min=0$ and $k_\max\approx 6$. 
These authors put forward that black holes are the source of dark energy, whereas we assume that dark energy can condense on
galaxies and end up in black holes.

When supermassive black holes are relevant for the mass budget, there will be an extra term in (\ref{Etot}).
Assuming for simplicity that $p(k)$ does not depend on $a$ and holds also for $a>1$, it takes the form
\BEQ
\Oma_\bh(a)=\Om_\bh\int\d k\,p(k) a^{k-3},
\EEQ
Black holes do not create pressure (Hawking radiation is negligible), 
so for $w_\bh=0$  the energy conservation (\refl{Friedmon}) yields an extra term to (\ref{Omla,Ea}),
\BEQ
\Om_\lam^\bh(a)=-\Om_\bh\int\d k \, p(k)k \frac{a^{k-3}-1}{k-3},
\EEQ
 vanishing at $a=1$, in accord with the definition $\Om_\Lam=\Om_\lam(1)$.
Eq. (\ref{Oma3=}) now reads 
\BEQ
\Oma(a)
=\frac{\Om_r}{a^4}+\frac{\Om_b }{a^3}+\frac{3a^{\delta_e-3}-\dele \, }{3-\dele}\Om_e +\Om_\Lam 
 +\Om_\bh\int\d k \, p(k) \frac{\, k-3a^{k-3}}{k-3},
\EEQ
with the closure $\Om_r+\Om_b+\Ome +\Om_\Lam+\Om_\bh=1$.
For large $a$ the integral behaves as $\sim - a^{k_\max-3} \ll - 1$, demonstrating that more energy is extracted 
from the aether than transferred to the black holes, due to the work cost to get it there.

Apparently, black holes have a tendency to make $\Oma$ negative, but since $E=H^2/H_0^2$, this is physically impossible.
If $\Oma$ reaches zero at some $a_\max$, this implies $\dot a=0$. Since values $\ddot a<0$ must then hold
already, it describes a Big Crunch with $\dot a<0$ from thereon.

\subsection{Times near the Big Bang}
\label{inflation}

Let us return to eq. (\ref{Friedmon}) and add a  term  $\Om_\phi(a)$ from an unspecified degree of freedom $\phi$,
\BEQ\label{Ebh(a)}
E(a)=\frac{\Om_r(a)}{a^4}+\frac{\Om_b(a)}{a^3}+\frac{\Ome (a)}{a^{3}}+\Om_\lam(a)+\Om_\phi(a),\quad 
\EEQ
Assuming that $\Om_r(a)$, $\Om_b(a)$, $\Ome(a) $ and $\Om_\phi(a)$ (but not $\Om_\lam$) all start at zero at $a=0$, 
and grow slowly enough, the quantity $\Om_\dS$, where dS stands for {\it de Sitter}, takes a finite value,
\BEQ\label{OmP=}
\Om_\dS=\Om_\Lam+\int_0^1 {\d b}\left [\frac{\Om_r'(b)}{b^4}+\frac{\Om_b'(b)}{b^3}+\frac{\Ome '(b)}{b^3} +3\frac{\Om_\phi(b)}{b}\right]+\Om_\phi(1),
\EEQ
Energy conservation (\ref{Friedmon}) allows to express $\Om_\lam$ with $\Om_\lam(1)=\Om_\Lam$ as
\BEQ \label{OmPding}
\Om_\lam(a) =\Om_\dS-\int_0^a {\d b}\left [\frac{\Om_r'(b)}{b^4}+\frac{\Om_b'(b)}{b^3}+\frac{\Ome '(b)}{b^3} 
+3(1+w_\phi)\frac{\Om_\phi(b)}{b}\right]  -\Om_\phi(a) .
\EEQ
After partial integration, eq.  (\ref{Ebh(a)})  reads
\BEQ \label{OmtotP}
E(a)=\Om_\dS-\int_0^a\frac{\d b}{b} \left[4\frac{\Om_r(b)}{b^4}+3\frac{\Om_b(b)}{b^3}+3\frac{\Ome (b)}{b^3} +3(1+w_\phi)\Om_\phi(b)\right] .
\EEQ
The decay of $E(a)$ is expressed here by the growth of the integral.

\subsubsection{High zero-point energy initial state: Automatic inflation}

The product $\rho_c\Om_\dS$ is the aether energy density at the big bang ($a\approx 0$). It may have the Planck value $\sim m_P^4$, 
so that $\Om_\dS\sim 10^{123}$, which is commonly seen as a catastrophic mismatch between theory and observation.
But, as mentioned in section \ref{zpe-physical},  that refers to the bare ZPE, which is unphysical; here it refers to the physical energy content
at the Big Bang.

Rather than doing away with the large ZPE, we  make it a cornerstone. We now assume that $\rho_c\Om_\dS=\rho_P\sim m_P^4$ 
is the physical zero point energy density injected in the quantum aether during the Big Bang, where $\Om_\dS\sim 10^{123}$.
This ZPE gets subsequently diluted by the expansion, by creating gravitational waves,  by turning it into particles and electrostatic energy, 
and by  participating in the dark matter.

The initial phase of the Universe is a de Sitter universe, with cosmological constant $H_\dS$ 
The Friedman equation ${\dot a^2}/{a^2}=H_0^2{E(a)}$ leads at early times, when only $E\approx \Om_\dS\gg1$ matters, 
to exponential expansion, $a(t)\approx  a_P\exp H_\dS t$ with  
$H_\dS=\sqrt{\Om_\dS} H_0$,  the inverse of the {\it  de Sitter time} 
$\tau_\dS=1/\sqrt{\Om_\dS}H_0$ $\approx 25 t_P$ for $\Om_P=10^{123}$ and $100t_P$ for $\Om_P=10^{120}$ 
with $t_P=\sqrt{\hbar G/c^5}=5.39\,10^{-44}$ s  the Planck time.
This behavior is called ``inflation''; in \EAE   theory, it happens automatically; an inflaton field $\phi$ seems needed to end the inflation.
In this classical approach, the initial time where $a=0$ appears to be $t_i=-\infty$.

In the course of time,  the subtraction terms in (\ref{OmPding}) and (\ref{OmtotP}) grow in size and diminish $E(a)$.
We assume that, after a certain period, these integrals creep towards $\Om_\dS$, leaving a relatively small $E(a)$,  and essentially 
make an end to the period of inflation. Such an end is generally expected to be smooth and called ``graceful exit'' (gx).

A candidate for enforcing the gx may be primordial black holes. It happens around  $a=a_\gx$ where
\BEQ
\Om_\dS-3\int_0^{a_\gx}\frac{\d b}{b}\Om_\phi(b) \ll \Om_\dS. 
\EEQ
From then on, the two terms in (\ref{OmtotP}) nearly cancel, and the expansion is more effectively expressed by the familiar form (\ref{Etot})
with $\Om_\phi(a)$ added, and with $\Om_\lam$ from (\ref{OmPding}) and (\ref{OmP=}) expressed in only moderately large terms, 
\BEQ
\Om_\lam=\Om_\Lam+\int_a^1 {\d b}\left [\frac{\Om_r'(b)}{b^4}+\frac{\Om_b'(b)}{b^3}+\frac{\Ome '(b)}{b^3} +3\frac{\Om_\phi(b)}{b}\right]
+\Om_\phi(1)-\Om_\phi(a),
\EEQ
In order not to ``overshoot'', i.e., not to make $E$ negative, $\Om_\phi(a)$ must have become relatively small near $a_\gx$.
For black holes, this may occur by Hawking evaporation, which leads to particle creation and thus increase of their temperature,
the so-called reheating.

\subsubsection{Low entropy initial state}

Roger Penrose has estimated the final entropy of the Universe by considering it as a huge black hole 
and applying the Bekenstein-Hawking formula. The result $S\sim 10^{123}$ \citep{roger1989emperor}
coincides with the above  $\Om_\dS\approx m_P^4/\rho_c$. This is much larger than the entropy $\sim10^{88}$ in the CMB radiation. 
Due to the second law, the entropy kept on increasing in the past, so it must even have been much smaller during the Big Bang.
The volume of phase space is $V=\exp S\sim \exp(10^{123})$, hence the question arises how the Creator could select our low entropy Universe
out of this enormous number of candidates \citep{roger1989emperor}.

In the above \EAE   initial state, only EZP/VE is present but no particles, neither photons nor gluons nor black holes, 
so that all field modes lie in their quantum ground state. Though general quantum systems are described by a mixed state, 
this case can be described by a pure quantum state, roughly as a product of individual ground states, like the Hartle-Hawking state \citep{hartle1983wave},
but with modified individual zero point energies to code the ZPE injection. 
The fine grained entropy, in this situation the von Neumann entropy, vanishes in a pure state.
Being conserved  under quantum dynamics, it vanishes at all times.
This puts many constraints on the ensuing dynamics, some of which are coded in known conservation laws.

Neglecting these correlations leads to a coarse grained entropy, that increases in time. 
Starting from zero, this solves Penrose's conundrum: in \EAE   theory a zero -- or at least, a low -- entropy initial state
is a consistency property, for which neither fine-tuning nor selection out of a vast set of candidate universes is involved.

\section{Conclusion}
\label{conclusion}

At the time of writing, there are two standard models. The first is the standard model of particle physics, formulated as 
a quantum field theory which is shown to be renormalizable by our teachers  Gerard 't Hooft and Tiny Veltman.
The second is $\Lambda$CDM, the standard model of cosmology, based on the assumptions of a cosmological constant and cold dark matter.
Next to the no-show in multiple CDM searches, this approach suffers from the Hubble tension.

Here we put forward that such a new type of matter is neither wanted nor needed, and
that standard models of cosmology and particle physics are actually one and the same.
No dark matter particle, which would require an extension of the standard model, 
is involved; a new view on the zero point energy of the vacuum suffices to explain
 the main constituents of the Universe, the 95\%  fraction of dark matter and dark energy.
Given that the Casimir effect for moving parallel conducting plates involves an inflow or outflow of zero point energy (ZPE),
we consider the ZPE as a fluid that can partly act as dark matter. Interpretation of our analytical results leads to consider the energy
of the vacuum itself as zero, or unmeasurable in any way,  while energy added to it, or taken out from it, 
acts as ``aether'' energy (though not the historic aether ruled out by the Michelson-Morley experiment), a physical component,
subject to the Einstein equations.

In this interpretation, we are led by the principle that one should first solve the mathematics and then provide a physical interpretation 
of the results, as applied to our approach to dynamics of quantum measurement and the ensuing statistical interpretation of quantum 
mechanics \citep{allahverdyan2013understanding,allahverdyan2017sub,allahverdyan2023teaching}.

The Einstein equations require that ZPE is assisted by an electric field, which can arise from a tiny mismatch between plus and minus charges in cosmic plasmas.
The combination is termed electro-zero-point energy (EZPE) or electro-aether-energy (EAE ), which aims to replace the popular cold dark matter.
In fact, the connection to an electric field seems natural but is not compulsary; its role may be taken by any vector field producing the structure
 $\rho_E\cC^\mu_{\ednu}$ in eq. (\ref{TMnrhomA}).

Rather than invoking new physics, \EAE   theory takes a new view on the capacities of the zero point energy of the (quantum) 
vacuum or just the energy of the classical vacuum.
We are led to view it not as a static, uniform entity but as a type of fluid, that can condense on mass concentrations.
This application of the standard model appoints an indispensable, and even leading role for the aether, 
to function as the main actor in cosmic structures by providing, presently, 70\% of the total mass/energy as dark energy
and 6\% involved as the ZPE part of the dark matter, combined with 19\% electrostatic energy in the dark matter.
Particles, in the form of normal matter, only play a secondary role, coming into existence later in the early Universe 
and forming presently some 5\% of the total mass.

\EAE   theory predicts that the dark matter present during the emission of cosmic background radiation arose from ZPE/AE condensation;
this leads naturally to the assumption of primordial black holes. They may have grown by \EAE   condensation and merging.
Black holes from stellar collapse can likewise grow by gentle inflow of \EAEc, filling ``mass gaps'' and triggering the growth of 
supermassive BHs not dominated by merging. Massive BHs  may ``steal'' the \EAE   from small surrounding ones, the tidal field effect.

Next, the BHs organize the galaxy around them, by an interdependent propensity for the available zero point energy,
that partly streams in from infinity and is partly taken out of the vacuum in the outskirts. 
To achieve this, the charge mismatch has to be optimal according to the Einstein-Coulomb equations.
In a galaxy and in a cluster there is a dynamical connection with the baryons:  
in order to host more ZPE coming in from infinity, an adjustment of the local net charge mismatch has to take place.
Flow of aether energy into black holes requires inflow of charges,  assuring a dynamical connection between the central BH and the whole galaxy.

A dynamical instability is identified, which assists in a speedy buildup of galactic and cluster cores with constant DM density,
supporting \EAE   theory and observations on the cusp-core problem.  The reverse  mechanism 
can explain the expansion and possible dissolution of galactic and cluster cores.

Hydrostatic equilibrium in galaxy clusters satisfied in \EAEd.
 
On cosmological scales the \EAE   acts as a pressureless type of cold dark matter. 
\EAE   theory goes even one step further: the ``cosmological constant'' measured from supernovae
is merely the present value of the dynamical zero point energy, that may have started out
at the Big Bang with the field theoretic value larger by some 123 orders of magnitude.
In \EAE   theory there is no fine-tuning, the  ``cosmological constant''  is small, 
since it is the present value of a decaying function.

Despite Einstein's most famous equation $E=mc^2$, \EAE   theory involves a discrepancy between mass and energy. 
Mass and matter are related to particles, including photons, while energy relates to a modified aether without further particles.
The kinetic energy of particles remains included in the ``mass''.

The analysis of the present  section shows that the final parsec problem is solved by mass accretion as it happens for \EAEd. 
Two supermassive BHs at parsec distance will finally merge by absorbing aether energy, which enhances the probability for observing gravitational waves 
from merging events by the future LISA system.

\section{Summary}
\label{summary}

Electro-aether energy (\EAE  ) relies on electric fields and the zero point energy of the quantum fields of the standard model
of particle physics. Alternatively, it is just a property of the ``classical'' vacuum.
The energy, often equated to a cosmological constant, actually gets depleted in its condensation as part of dark matter.
These insights explain a cornucopia of  phenomena.

After considering various aspects of ZPE in section 2, the \EAE   framework is laid out in sec. 3. 
For spherically symmetric setups, it is shown how a non-uniform ZPE, combined with an electric field, is compatible with the Einstein equations.
ZPE is absorbed from the environment, while subject to a reshuffling inside the galaxy or cluster; 
its density is positive inside a core region and negative in the halo region.
There results a core with a net plus charge, surrounded by a halo with net minus charge; the total charge is zero.

A stability analysis is carried out. 
An inhomogeneous solution is connected to the formation of dark matter cores made up of \EAEc, and their later dissolution.
 
Estimates for various quantities are discussed in section 4. Particular attention is paid to the net charge fraction in the plasma.
While standard estimates allow maximally a fraction of $10^{-18}$, \EAE   involves a fraction that can perhaps be $10^5$ times larger.
Analysis of the hydrostatic equilibrium shows that the strong Coulomb repulsion and attraction is counteracted 
by the negative casu quo positive gradient of the ZPE pressure.

Section 5 deals with black holes. They can grow by \EAE   accretion, which rules out the ``mass gaps'' from standard arguments, 
and is supported by some gravitational wave events. 
It is shown that the final parsec problem for black hole merging is overcome 
by ongoing mass accretion within \EAE   theory.

Section 6 considered the application to galaxies. It is postulated that results from Modified Newtonian Dynamics (MOND)
can be modeled by \EAE   theory, and that the involved electric field and underlying charge density regulates
a connection between the dark matter structure, the shape of the rotation curve and the central supermassive black hole.
It is pointed out that constant-density dark matter cores, more than cusped ones, should be expected, and support for this is reviewed. 

In the application to galaxy clusters of section 7, first an isothermal sphere-type of fit is worked out for strong lensing data
of two fat clusters and the relation to the charge distributions and ZPE profiles is worked out.
Special attention is payed to their hydrodynamic equilibrium puzzle,  solved in \EAE  theory.

For the application to cosmology, section 8  first shows that at cosmological scales, the pressure connected to \EAE   dark matter
vanishes, as desired.
It is pointed out that ongoing ZPE condensation leads to a late-time increasing amount of dark matter. 
A fit of the $\Lam$CDM theory for the Planck data for the cosmic microwave background already softens the Hubble tension
between its value $H_0=68$ km/s Mpc and the late-time value $H_0=73$ km/s Mpc from supernovae. 
To investigate a full resolution of the problem, the CMB theory within \EAE   theory needs to be worked out and fitted.
Next, the cosmologically-near and far future is considered. 
A big crunch scenario is worked out involving a dominant role of black holes.
Finally,  attention is paid to the Big Bang period, 
where it is assumed that a large cosmological ``constant'' is inserted, leading to an  initial state with large aether energy and zero entropy.
Inflation occurs automatically.

Various further dark matter aspects in galaxies, clusters and cosmology seem to fall into place like pieces of a jigsaw puzzle,
see Section 9 of ref. \citep{nieuwenhuizen2023solution}.

\section{Outlook}
\label{outlook}

In  \EAE   cosmology, there is no dark matter particle, as supported by the no-show in dark matter searches,
but the theory gets ruled out when such a detection is made.

Simulations for the \EAE   paradigm are desired to test it on various observations,  replacing the current $\Lam$CDM simulations.
Given the great expertise in the latter, the situation seems hopeful. Irrespective of our proposal,
the recent James Webb  Space Telescope observation of very early onset of massive galaxy formation \citep{labbe2022very}
already seems to demand a new understanding of structure formation.

With the arrival of a new standard model, many issues in cosmology may hope for explanation.
We have mentioned the Hubble tension, softened already,  the Lithium-7 problem and hinted at black holes 
for a big crunch and, perhaps,  a gentle exit of inflation.

The predicted smaller age of the Universe of  some 12.8 Gyr poses questions regarding the earliest stars and structures;
these are not new, however,  since they follow mainly from adopting the large value of the present Hubble constant.
In this regard, early black hole growth by \EAE   accretion and  early galaxy formation due to rupture of hydrogen clouds 
may emerge as a consistent picture, allowing vast polar structures of matter around them  due to the \EAE   scaffold.
Being charged locally, the expanding primordial  hydrogen cloud will be subject to lightnings, after which the thunders
may provide the onset of prolate and/or barred structure formation.

The question ``why is the cosmological constant so small'' gets the \EAE   answer: 
the cosmological ``constant'' depends on space and time; 
there is no fine-tuning,  during the Big Bang a large zero point energy was inserted, which decreased.
It was a cold Big Bang, that quickly became hot.
The possibility for a recollapsing Universe may relate to a cyclic repetition of expansions and collapses.

In all these processes, aether energy (vacuum energy) is the ideal servant, an obedient, malleable agency, doing just the right thing  at the right time.
One may wonder whether it plays a similar role in standard, terrestrial electrostatic and perhaps magnetostatic problems.
Progress in this direction will be reported elsewhere \citep{nieuwenhuizen2024aether}.
Lastly, one may wonder whether, as in black holes, also in cosmology, the rotation of structures can carry some of the burden of the net charges.

\section{Acknowledgements}

It is a pleasure to thank Rudolf Sprik, Ben van Linden van den Heuvell, Ralph Wijers,  Indranil Banik,
Piet Mulders, Peter Keefe, Rudy Schild and Jasper van Wezel for discussion.


\begin{thebibliography}{126}
\providecommand{\natexlab}[1]{#1}
\expandafter\ifx\csname urlstyle\endcsname\relax
  \providecommand{\doi}[1]{doi:\discretionary{}{}{}#1}\else
  \providecommand{\doi}{doi:\discretionary{}{}{}\begingroup
  \urlstyle{rm}\Url}\fi
\providecommand{\selectlanguage}[1]{\relax}
\providecommand{\bibAnnoteFile}[1]{%
  \IfFileExists{#1}{\begin{quotation}\noindent\textsc{Key:} #1\\
  \textsc{Annotation:}\ \input{#1}\end{quotation}}{}}
\providecommand{\bibAnnote}[2]{%
  \begin{quotation}\noindent\textsc{Key:} #1\\
  \textsc{Annotation:}\ #2\end{quotation}}

\bibitem[{Abbott et~al.(2024)Abbott, Acevedo, Aguena, Alarcon, Allam, Alves
  et~al.}]{abbott2024dark}
Abbott, T., Acevedo, M., Aguena, M., Alarcon, A., Allam, S., Alves, O., et~al.
  (2024).
\newblock The {D}ark {E}nergy {S}urvey: {C}osmology results with $\sim1500$ new
  high-redshift type ia supernovae using the full 5-year dataset.
\newblock \emph{arXiv preprint arXiv:2401.02929}
\input{abbott2024dark}

\bibitem[{Aghanim et~al.(2020)Aghanim, Akrami, Ashdown, Aumont, Baccigalupi,
  Ballardini et~al.}]{aghanim2020planck}
Aghanim, N., Akrami, Y., Ashdown, M., Aumont, J., Baccigalupi, C., Ballardini,
  M., et~al. (2020).
\newblock Planck 2018 results-vi. cosmological parameters.
\newblock \emph{Astronomy \& Astrophysics} 641, A6
\input{aghanim2020planck}

\bibitem[{Alcock et~al.(2000)Alcock, Allsman, Alves, Axelrod, Becker, Bennett
  et~al.}]{alcock2000macho}
Alcock, C., Allsman, R., Alves, D.~R., Axelrod, T., Becker, A.~C., Bennett, D.,
  et~al. (2000).
\newblock The macho project: microlensing results from 5.7 years of {L}arge
  {M}agellanic {C}loud observations.
\newblock \emph{The Astrophysical Journal} 542, 281
\input{alcock2000macho}

\bibitem[{Allahverdyan et~al.(2013)Allahverdyan, Balian, and
  Nieuwenhuizen}]{allahverdyan2013understanding}
Allahverdyan, A.~E., Balian, R., and Nieuwenhuizen, T.~M. (2013).
\newblock Understanding quantum measurement from the solution of dynamical
  models.
\newblock \emph{Physics Reports} 525, 1--166
\input{allahverdyan2013understanding}

\bibitem[{Allahverdyan et~al.(2017)Allahverdyan, Balian, and
  Nieuwenhuizen}]{allahverdyan2017sub}
Allahverdyan, A.~E., Balian, R., and Nieuwenhuizen, T.~M. (2017).
\newblock A sub-ensemble theory of ideal quantum measurement processes.
\newblock \emph{Annals of Physics} 376, 324--352
\input{allahverdyan2017sub}

\bibitem[{Allahverdyan et~al.(2024)Allahverdyan, Balian, and
  Nieuwenhuizen}]{allahverdyan2023teaching}
Allahverdyan, A.~E., Balian, R., and Nieuwenhuizen, T.~M. (2024).
\newblock Teaching ideal quantum measurement, from dynamics to interpretation.
\newblock \emph{Comptes Rendus Physique}
\input{allahverdyan2023teaching}

\bibitem[{Arcadi et~al.(2018)Arcadi, Dutra, Ghosh, Lindner, Mambrini, Pierre
  et~al.}]{arcadi2018waning}
Arcadi, G., Dutra, M., Ghosh, P., Lindner, M., Mambrini, Y., Pierre, M., et~al.
  (2018).
\newblock The waning of the {WIMP}? a {R}eview of models, searches, and
  constraints.
\newblock \emph{The European Physical Journal C} 78, 1--57
\input{arcadi2018waning}

\bibitem[{Arita et~al.(2023)Arita, Kashikawa, Matsuoka, He, Ito, Liang
  et~al.}]{arita2023subaru}
Arita, J., Kashikawa, N., Matsuoka, Y., He, W., Ito, K., Liang, Y., et~al.
  (2023).
\newblock Subaru high-$z$ exploration of low-luminosity quasars ({SHELLQ}s).
  {XVIII}. {T}he dark matter halo mass of quasars at $z\sim 6$.
\newblock \emph{The Astrophysical Journal} 954, 210
\input{arita2023subaru}

\bibitem[{Armitage and Natarajan(2002)}]{armitage2002accretion}
Armitage, P.~J. and Natarajan, P. (2002).
\newblock Accretion during the merger of supermassive black holes.
\newblock \emph{The Astrophysical Journal} 567, L9
\input{armitage2002accretion}

\bibitem[{Bais and Russell(1975)}]{bais1975magnetic}
Bais, F. and Russell, R. (1975).
\newblock Magnetic-monopole solution of non-{A}belian gauge theory in curved
  spacetime.
\newblock \emph{Physical {R}eview D} 11, 2692
\input{bais1975magnetic}

\bibitem[{Balian and Duplantier(2004)}]{balian2004geometry}
Balian, R. and Duplantier, B. (2004).
\newblock Geometry of the {C}asimir effect.
\newblock \emph{arXiv preprint quant-ph/0408124}
\input{balian2004geometry}

\bibitem[{Ba{\~n}ados et~al.(2018)Ba{\~n}ados, Venemans, Mazzucchelli, Farina,
  Walter, Wang et~al.}]{banados2018800}
Ba{\~n}ados, E., Venemans, B.~P., Mazzucchelli, C., Farina, E.~P., Walter, F.,
  Wang, F., et~al. (2018).
\newblock An 800-million-solar-mass black hole in a significantly neutral
  {U}niverse at a redshift of 7.5.
\newblock \emph{Nature} 553, 473--476
\input{banados2018800}

\bibitem[{Banik et~al.(2024)Banik, Pittordis, Sutherland, Famaey, Ibata, Mieske
  et~al.}]{banik2024strong}
Banik, I., Pittordis, C., Sutherland, W., Famaey, B., Ibata, R., Mieske, S.,
  et~al. (2024).
\newblock Strong constraints on the gravitational law from {G}aia {DR}3 wide
  binaries.
\newblock \emph{Monthly Notices of the Royal Astronomical Society} 527,
  4573--4615
\input{banik2024strong}

\bibitem[{Banik and Zhao(2022)}]{banik2022galactic}
Banik, I. and Zhao, H. (2022).
\newblock From galactic bars to the {H}ubble tension: {W}eighing up the
  astrophysical evidence for {M}ilgromian gravity.
\newblock \emph{Symmetry} 14, 1331
\input{banik2022galactic}

\bibitem[{Bartels et~al.(2018)Bartels, Calore, Storm, and
  Weniger}]{bartels2018galactic}
Bartels, R., Calore, F., Storm, E., and Weniger, C. (2018).
\newblock Galactic binaries can explain the {F}ermi {G}alactic centre excess
  and 511 ke{V} emission.
\newblock \emph{Monthly Notices of the Royal Astronomical Society} 480,
  3826--3841
\input{bartels2018galactic}

\bibitem[{Beck et~al.(2013)Beck, Balogh, Bykov, Treumann, and
  Widrow}]{beck2013large}
Beck, R., Balogh, A., Bykov, A., Treumann, R.~A., and Widrow, L.~M. (2013).
\newblock \emph{Large-Scale Magnetic Fields in the {U}niverse} (Springer)
\input{beck2013large}

\bibitem[{Bertone(2010)}]{bertone2010particle}
Bertone, G. (2010).
\newblock \emph{Particle dark matter: observations, models and searches}
  (Cambridge University Press)
\input{bertone2010particle}

\bibitem[{Blas et~al.(2011)Blas, Lesgourgues, and Tram}]{blas2011cosmic}
Blas, D., Lesgourgues, J., and Tram, T. (2011).
\newblock The cosmic linear anisotropy solving system (class). {P}art {II}:
  approximation schemes.
\newblock \emph{Journal of Cosmology and Astroparticle Physics} 2011, 034
\input{blas2011cosmic}

\bibitem[{Blumenthal et~al.(1984)Blumenthal, Faber, Primack, and
  Rees}]{blumenthal1984formation}
Blumenthal, G.~R., Faber, S., Primack, J.~R., and Rees, M.~J. (1984).
\newblock Formation of galaxies and large-scale structure with cold dark
  matter.
\newblock \emph{Nature} 311, 517--525
\input{blumenthal1984formation}

\bibitem[{Blumenthal et~al.(1982)Blumenthal, Pagels, and
  Primack}]{blumenthal1982galaxy}
Blumenthal, G.~R., Pagels, H., and Primack, J.~R. (1982).
\newblock Galaxy formation by dissipationless particles heavier than neutrinos.
\newblock \emph{Nature} 299, 37--38
\input{blumenthal1982galaxy}

\bibitem[{Boldrini(2021)}]{boldrini2021cusp}
Boldrini, P. (2021).
\newblock The cusp--core problem in gas-poor dwarf spheroidal galaxies.
\newblock \emph{Galaxies} 10, 5
\input{boldrini2021cusp}

\bibitem[{Bond et~al.(1982)Bond, Szalay, and Turner}]{bond1982formation}
Bond, J.~R., Szalay, A.~S., and Turner, M.~S. (1982).
\newblock Formation of galaxies in a gravitino-dominated {U}niverse.
\newblock \emph{Physical {R}eview {L}etters} 48, 1636
\input{bond1982formation}

\bibitem[{Boyarsky et~al.(2014)Boyarsky, Ruchayskiy, Iakubovskyi, and
  Franse}]{boyarsky2014unidentified}
Boyarsky, A., Ruchayskiy, O., Iakubovskyi, D., and Franse, J. (2014).
\newblock Unidentified line in {X}-ray spectra of the {A}ndromeda galaxy and
  {P}erseus galaxy cluster.
\newblock \emph{Physical {R}eview {L}etters} 113, 251301
\input{boyarsky2014unidentified}

\bibitem[{Boyer(1968)}]{boyer1968quantum}
Boyer, T.~H. (1968).
\newblock Quantum electromagnetic zero-point energy of a conducting spherical
  shell and the {C}asimir model for a charged particle.
\newblock \emph{Physical {R}eview} 174, 1764
\input{boyer1968quantum}

\bibitem[{Brout et~al.(2022)Brout, Scolnic, Popovic, Riess, Carr, Zuntz
  et~al.}]{brout2022pantheon}
Brout, D., Scolnic, D., Popovic, B., Riess, A.~G., Carr, A., Zuntz, J., et~al.
  (2022).
\newblock The {P}antheon+ analysis: cosmological constraints.
\newblock \emph{The Astrophysical Journal} 938, 110
\input{brout2022pantheon}

\bibitem[{Brouwer et~al.(2017)Brouwer, Visser, Dvornik, Hoekstra, Kuijken,
  Valentijn et~al.}]{brouwer2017first}
Brouwer, M.~M., Visser, M.~R., Dvornik, A., Hoekstra, H., Kuijken, K.,
  Valentijn, E.~A., et~al. (2017).
\newblock First test of {V}erlinde's theory of emergent gravity using weak
  {G}ravitational lensing measurements.
\newblock \emph{Monthly Notices of the Royal Astronomical Society} 466,
  2547--2559
\input{brouwer2017first}

\bibitem[{Bull et~al.(2016)Bull, Akrami, Adamek, Baker, Bellini, Jim{\'e}nez
  et~al.}]{bull2016beyond}
Bull, P., Akrami, Y., Adamek, J., Baker, T., Bellini, E., Jim{\'e}nez, J.~B.,
  et~al. (2016).
\newblock Beyond $\lambda${CDM}: {P}roblems, solutions, and the road ahead.
\newblock \emph{Physics of the Dark {U}niverse} 12, 56--99
\input{bull2016beyond}

\bibitem[{Bullock and Boylan-Kolchin(2017)}]{bullock2017small}
Bullock, J.~S. and Boylan-Kolchin, M. (2017).
\newblock Small-scale challenges to the $\lambda${CDM} paradigm.
\newblock \emph{Annu. Rev. Astron. Astrophys} 55, 343--87
\input{bullock2017small}

\bibitem[{Camarena and Marra(2020{\natexlab{a}})}]{camarena2020local}
Camarena, D. and Marra, V. (2020{\natexlab{a}}).
\newblock Local determination of the {H}ubble constant and the deceleration
  parameter.
\newblock \emph{Physical {R}eview Research} 2, 013028
\input{camarena2020local}

\bibitem[{Camarena and Marra(2020{\natexlab{b}})}]{camarena2020new}
Camarena, D. and Marra, V. (2020{\natexlab{b}}).
\newblock A new method to build the (inverse) distance ladder.
\newblock \emph{Monthly Notices of the Royal Astronomical Society} 495,
  2630--2644
\input{camarena2020new}

\bibitem[{{C}asimir(1948)}]{casimir1948attraction}
{C}asimir, H.~B. (1948).
\newblock On the attraction between two perfectly conducting plates.
\newblock In \emph{Proc. Kon. Ned. Akad. Wet.} vol.~51, 793
\input{casimir1948attraction}

\bibitem[{Cetto and De~La Pe\~na(1993)}]{cetto1993casimir}
Cetto, A. and De~La Pe\~na, L. (1993).
\newblock {C}asimir effect for bodies of arbitrary size.
\newblock \emph{Il Nuovo Cimento B (1971-1996)} 108, 447--458
\input{cetto1993casimir}

\bibitem[{Chakraborty et~al.(2014)Chakraborty, Rahaman, Ray, Nandi, and
  Islam}]{chakraborty2014possible}
Chakraborty, K., Rahaman, F., Ray, S., Nandi, A., and Islam, N. (2014).
\newblock Possible features of galactic halo with electric field and
  observational constraints.
\newblock \emph{General Relativity and {G}ravitation} 46, 1--24
\input{chakraborty2014possible}

\bibitem[{Coc and Vangioni(2005)}]{coc2005lithium}
Coc, A. and Vangioni, E. (2005).
\newblock Lithium and {B}ig-{B}ang {N}ucleosynthesis.
\newblock \emph{Proceedings of the International Astronomical Union} 1, 13--22
\input{coc2005lithium}

\bibitem[{Collaboration et~al.(2022)}]{lhcb2022measurement}
Collaboration, L. et~al. (2022).
\newblock Measurement of lepton universality parameters in
  {B}$^+\mapsto${K}$^+\ell^+\ell^-$ and {B}$^0\mapsto${K}$^{\ast
  0}\ell^+\ell^-$ decays.
\newblock \emph{arXiv preprint arXiv:2212.09153}
\input{lhcb2022measurement}

\bibitem[{Cook(2001)}]{cook2001hendrik}
Cook, A. (2001).
\newblock Hendrik {C}hristoffel van de {H}ulst {R}idder in de {O}rde van
  {N}ederlandse {L}eeuw. 19 {N}ovember 1918-31 {J}uly 2000.
\newblock \emph{Biographical Memoirs of Fellows of the Royal Society} 47,
  467--479
\input{cook2001hendrik}

\bibitem[{Corda(2009)}]{corda2009interferometric}
Corda, C. (2009).
\newblock Interferometric detection of gravitational waves: the definitive test
  for general relativity.
\newblock \emph{International Journal of Modern Physics D} 18, 2275--2282
\input{corda2009interferometric}

\bibitem[{Creevey et~al.(2015)Creevey, Th{\'e}venin, Berio, Heiter, Von~Braun,
  Mourard et~al.}]{creevey2015benchmark}
Creevey, O., Th{\'e}venin, F., Berio, P., Heiter, U., Von~Braun, K., Mourard,
  D., et~al. (2015).
\newblock Benchmark stars for {G}aia {F}undamental properties of the
  {P}opulation ii star {H}{D} 140283 from interferometric, spectroscopic, and
  photometric data.
\newblock \emph{Astronomy \& Astrophysics} 575, A26
\input{creevey2015benchmark}

\bibitem[{Curtis-Lake et~al.(2023)Curtis-Lake, Carniani, Cameron, Charlot,
  Jakobsen, Maiolino et~al.}]{curtis2023spectroscopic}
Curtis-Lake, E., Carniani, S., Cameron, A., Charlot, S., Jakobsen, P.,
  Maiolino, R., et~al. (2023).
\newblock Spectroscopic confirmation of four metal-poor galaxies at $z=$
  10.3--13.2.
\newblock \emph{Nature Astronomy} 7, 622--632
\input{curtis2023spectroscopic}

\bibitem[{Dainotti et~al.(2021)Dainotti, De~Simone, Schiavone, Montani,
  Rinaldi, and Lambiase}]{dainotti2021hubble}
Dainotti, M.~G., De~Simone, B., Schiavone, T., Montani, G., Rinaldi, E., and
  Lambiase, G. (2021).
\newblock On the {H}ubble constant tension in the {SN}e ia {P}antheon sample.
\newblock \emph{The Astrophysical Journal} 912, 150
\input{dainotti2021hubble}

\bibitem[{De~Blok et~al.(2001)De~Blok, McGaugh, Bosma, and
  Rubin}]{deblok2001mass}
De~Blok, W., McGaugh, S.~S., Bosma, A., and Rubin, V.~C. (2001).
\newblock Mass density profiles of low surface brightness galaxies.
\newblock \emph{The Astrophysical Journal} 552, L23
\input{deblok2001mass}

\bibitem[{De~Swart et~al.(2017)De~Swart, Bertone, and van Dongen}]{de2017dark}
De~Swart, J., Bertone, G., and van Dongen, J. (2017).
\newblock How dark matter came to matter.
\newblock \emph{Nature Astronomy} 1, 1--9
\input{de2017dark}

\bibitem[{Del~Popolo and Le~Delliou(2021)}]{del2021review}
Del~Popolo, A. and Le~Delliou, M. (2021).
\newblock {R}eview of solutions to the cusp-core problem of the $\lambda${CDM}
  model.
\newblock \emph{Galaxies} 9, 123
\input{del2021review}

\bibitem[{Di~Dio et~al.(2013)Di~Dio, Montanari, Lesgourgues, and
  Durrer}]{di2013classgal}
Di~Dio, E., Montanari, F., Lesgourgues, J., and Durrer, R. (2013).
\newblock The {CLASS}gal code for relativistic cosmological large scale
  structure.
\newblock \emph{Journal of Cosmology and Astroparticle Physics} 2013, 044
\input{di2013classgal}

\bibitem[{Di~Paolo et~al.(2019)Di~Paolo, Salucci, and Erkurt}]{di2019universal}
Di~Paolo, C., Salucci, P., and Erkurt, A. (2019).
\newblock The universal rotation curve of low surface brightness
  galaxies--{IV}. {T}he interrelation between dark and luminous matter.
\newblock \emph{Monthly Notices of the Royal Astronomical Society} 490,
  5451--5477
\input{di2019universal}

\bibitem[{Dreitlein(1974)}]{dreitlein1974broken}
Dreitlein, J. (1974).
\newblock Broken symmetry and the cosmological constant.
\newblock \emph{Physical {R}eview {L}etters} 33, 1243
\input{dreitlein1974broken}

\bibitem[{Farrah et~al.(2023{\natexlab{a}})Farrah, Croker, Zevin, Tarl{\'e},
  Faraoni, Petty et~al.}]{farrah2023observational}
Farrah, D., Croker, K.~S., Zevin, M., Tarl{\'e}, G., Faraoni, V., Petty, S.,
  et~al. (2023{\natexlab{a}}).
\newblock Observational evidence for cosmological coupling of black holes and
  its implications for an astrophysical source of dark energy.
\newblock \emph{The Astrophysical Journal {L}etters} 944, L31
\input{farrah2023observational}

\bibitem[{Farrah et~al.(2023{\natexlab{b}})Farrah, Petty, Croker, Tarl{\'e},
  Zevin, Hatziminaoglou et~al.}]{farrah2023preferential}
Farrah, D., Petty, S., Croker, K.~S., Tarl{\'e}, G., Zevin, M., Hatziminaoglou,
  E., et~al. (2023{\natexlab{b}}).
\newblock A preferential growth channel for supermassive black holes in
  elliptical galaxies at $z\le 2$.
\newblock \emph{The Astrophysical Journal} 943, 133
\input{farrah2023preferential}

\bibitem[{Gurnett et~al.(2013)Gurnett, Kurth, Burlaga, and
  Ness}]{gurnett2013situ}
Gurnett, D., Kurth, W., Burlaga, L., and Ness, N. (2013).
\newblock In situ observations of interstellar plasma with {V}oyager 1.
\newblock \emph{Science} 341, 1489--1492
\input{gurnett2013situ}

\bibitem[{Hartle and Hawking(1983)}]{hartle1983wave}
Hartle, J.~B. and Hawking, S.~W. (1983).
\newblock Wave function of the {U}niverse.
\newblock \emph{Physical {R}eview D} 28, 2960
\input{hartle1983wave}

\bibitem[{Hashimoto et~al.(2023)Hashimoto, {\'A}lvarez-M{\'a}rquez, Fudamoto,
  Colina, Inoue, Nakazato et~al.}]{hashimoto2023reionization}
Hashimoto, T., {\'A}lvarez-M{\'a}rquez, J., Fudamoto, Y., Colina, L., Inoue,
  A., Nakazato, Y., et~al. (2023).
\newblock Reionization and the {ISM}/stellar origins with {{JWST}} and {ALMA}
  ({RIOJA}): {T}he core of the highest-redshift galaxy overdensity at $z= 7.88$
  confirmed by {NIRS}pec/{{JWST}}.
\newblock \emph{The Astrophysical Journal {L}etters} 955, L2
\input{hashimoto2023reionization}

\bibitem[{Kafle et~al.(2014)Kafle, Sharma, Lewis, and
  Bland-Hawthorn}]{kafle2014shoulders}
Kafle, P.~R., Sharma, S., Lewis, G.~F., and Bland-Hawthorn, J. (2014).
\newblock On the shoulders of giants: properties of the stellar halo and the
  {M}ilky {W}ay mass distribution.
\newblock \emph{The Astrophysical Journal} 794, 59
\input{kafle2014shoulders}

\bibitem[{Kapteyn(2013)}]{kapteyn2013First}
Kapteyn, J.~C. (2013).
\newblock First {A}ttempt at a {T}heory of the {A}rrangement and {M}otion of
  the {S}idereal {S}ystem.
\newblock In \emph{A Source Book in Astronomy and Astrophysics, 1900--1975}
  (Harvard University Press). 542--549
\input{kapteyn2013First}

\bibitem[{Karukes and Salucci(2017)}]{karukes2017universal}
Karukes, E.~V. and Salucci, P. (2017).
\newblock The universal rotation curve of dwarf disc galaxies.
\newblock \emph{Monthly Notices of the Royal Astronomical Society} 465,
  4703--4722
\input{karukes2017universal}

\bibitem[{Keeley and Shafieloo(2023)}]{keeley2023ruling}
Keeley, R.~E. and Shafieloo, A. (2023).
\newblock Ruling out new physics at low redshift as a solution to the $h_0$
  tension.
\newblock \emph{Physical {R}eview {L}etters} 131, 111002
\input{keeley2023ruling}

\bibitem[{Kroupa(2012)}]{kroupa2012dark}
Kroupa, P. (2012).
\newblock The dark matter crisis: falsification of the current standard model
  of cosmology.
\newblock \emph{Publications of the Astronomical Society of Australia} 29,
  395--433
\input{kroupa2012dark}

\bibitem[{Kroupa(2014)}]{kroupa2014planar}
Kroupa, P. (2014).
\newblock The planar satellite distributions around {A}ndromeda, the {M}ilky
  {W}ay and other galaxies, and their implications for fundamental physics.
\newblock \emph{Multi-Spin Galaxies} 486, 183
\input{kroupa2014planar}

\bibitem[{Kroupa(2015)}]{kroupa2015galaxies}
Kroupa, P. (2015).
\newblock Galaxies as simple dynamical systems: observational data disfavor
  dark matter and stochastic star formation.
\newblock \emph{Canadian Journal of Physics} 93, 169--202
\input{kroupa2015galaxies}

\bibitem[{Labb\'e et~al.(2022)Labb\'e, van Dokkum, Nelson, Bezanson, Suess,
  Leja et~al.}]{labbe2022very}
Labb\'e, I., van Dokkum, P., Nelson, E., Bezanson, R., Suess, K., Leja, J.,
  et~al. (2022).
\newblock A very early onset of massive galaxy formation.
\newblock \emph{arXiv preprint arXiv:2207.12446}
\input{labbe2022very}

\bibitem[{Land and Magueijo(2005)}]{land2005examination}
Land, K. and Magueijo, J. (2005).
\newblock Examination of evidence for a preferred axis in the cosmic radiation
  anisotropy.
\newblock \emph{Physical Review Letters} 95, 071301
\input{land2005examination}

\bibitem[{Lelli et~al.(2017)Lelli, McGaugh, Schombert, and
  Pawlowski}]{lelli2017one}
Lelli, F., McGaugh, S.~S., Schombert, J.~M., and Pawlowski, M.~S. (2017).
\newblock One law to rule them all: the radial acceleration relation of
  galaxies.
\newblock \emph{The Astrophysical Journal} 836, 152
\input{lelli2017one}

\bibitem[{Lemze et~al.(2011)Lemze, Rephaeli, Barkana, Broadhurst, Wagner, and
  Norman}]{lemze2011quantifying}
Lemze, D., Rephaeli, Y., Barkana, R., Broadhurst, T., Wagner, R., and Norman,
  M.~L. (2011).
\newblock Quantifying the collisionless nature of dark matter and galaxies in
  {A}1689.
\newblock \emph{The Astrophysical Journal} 728, 40
\input{lemze2011quantifying}

\bibitem[{Linde(1974)}]{linde1974lee}
Linde, A.~D. (1974).
\newblock Is the {L}ee constant a cosmological constant.
\newblock \emph{JETP Lett} 19, 183
\input{linde1974lee}

\bibitem[{L{\'o}pez~Fune et~al.(2017)L{\'o}pez~Fune, Salucci, and
  Corbelli}]{lopez2017radial}
L{\'o}pez~Fune, E., Salucci, P., and Corbelli, E. (2017).
\newblock Radial dependence of the dark matter distribution in {M}33.
\newblock \emph{Monthly Notices of the Royal Astronomical Society} 468,
  147--153
\input{lopez2017radial}

\bibitem[{Maiolino et~al.(2024)Maiolino, Scholtz, Witstok, Carniani,
  D’Eugenio, de~Graaff et~al.}]{maiolino2024small}
Maiolino, R., Scholtz, J., Witstok, J., Carniani, S., D’Eugenio, F.,
  de~Graaff, A., et~al. (2024).
\newblock A small and vigorous black hole in the early {U}niverse.
\newblock \emph{Nature} , 1--3
\input{maiolino2024small}

\bibitem[{Metz and Kroupa(2007)}]{metz2007dwarf}
Metz, M. and Kroupa, P. (2007).
\newblock Dwarf spheroidal satellites: are they of tidal origin?
\newblock \emph{Monthly Notices of the Royal Astronomical Society} 376,
  387--392
\input{metz2007dwarf}

\bibitem[{Migkas et~al.(2020)Migkas, Schellenberger, Reiprich, Pacaud,
  Ramos-Ceja, and Lovisari}]{migkas2020probing}
Migkas, K., Schellenberger, G., Reiprich, T., Pacaud, F., Ramos-Ceja, M., and
  Lovisari, L. (2020).
\newblock Probing cosmic isotropy with a new x-ray galaxy cluster sample
  through the lx--t scaling relation.
\newblock \emph{Astronomy \& Astrophysics} 636, A15
\input{migkas2020probing}

\bibitem[{Milgrom(1983)}]{milgrom1983modification}
Milgrom, M. (1983).
\newblock A modification of the {N}ewtonian dynamics as a possible alternative
  to the hidden mass hypothesis.
\newblock \emph{The Astrophysical Journal} 270, 365--370
\input{milgrom1983modification}

\bibitem[{Milosavljevi{\'c} and Merritt(2003)}]{milosavljevic2003final}
Milosavljevi{\'c}, M. and Merritt, D. (2003).
\newblock The final parsec problem.
\newblock In \emph{AIP Conference Proceedings} (American Institute of Physics),
  vol. 686, 201--210
\input{milosavljevic2003final}

\bibitem[{Milton(2003)}]{milton2003casimir}
Milton, K.~A. (2003).
\newblock \emph{The {C}asimir effect: physical manifestations of zero-point
  energy} (American Association of Physics Teachers)
\input{milton2003casimir}

\bibitem[{Molnar et~al.(2010)Molnar, Chiu, Umetsu, Chen, Hearn, Broadhurst
  et~al.}]{molnar2010testing}
Molnar, S., Chiu, I.-N., Umetsu, K., Chen, P., Hearn, N., Broadhurst, T.,
  et~al. (2010).
\newblock Testing strict hydrostatic equilibrium in simulated clusters of
  galaxies: implications for {A}1689.
\newblock \emph{The Astrophysical Journal {L}etters} 724, L1
\input{molnar2010testing}

\bibitem[{Morandi et~al.(2012)Morandi, Limousin, Sayers, Golwala, Czakon,
  Pierpaoli et~al.}]{morandi2012x}
Morandi, A., Limousin, M., Sayers, J., Golwala, S.~R., Czakon, N.~G.,
  Pierpaoli, E., et~al. (2012).
\newblock X-ray, lensing and {S}unyaev-{Z}el'{d}ovich triaxial analysis of
  {A}bell 1835 out to {R} 200.
\newblock \emph{Monthly Notices of the Royal Astronomical Society} 425,
  2069--2082
\input{morandi2012x}

\bibitem[{Morandi et~al.(2010)Morandi, Pedersen, and
  Limousin}]{morandi2010unveiling}
Morandi, A., Pedersen, K., and Limousin, M. (2010).
\newblock Unveiling the three-dimensional structure of galaxy clusters:
  resolving the discrepancy between {X}-ray and lensing masses.
\newblock \emph{The Astrophysical Journal} 713, 491
\input{morandi2010unveiling}

\bibitem[{Nanayakkara et~al.(2024)Nanayakkara, Glazebrook, Jacobs,
  Kawinwanichakij, Schreiber, Brammer et~al.}]{nanayakkara2024population}
Nanayakkara, T., Glazebrook, K., Jacobs, C., Kawinwanichakij, L., Schreiber,
  C., Brammer, G., et~al. (2024).
\newblock A population of faint, old, and massive quiescent galaxies at $3< z<
  4$ revealed by {JWST} {NIRS}pec spectroscopy.
\newblock \emph{Scientific Reports} 14, 3724
\input{nanayakkara2024population}

\bibitem[{Navarro et~al.(1997)Navarro, Frenk, and White}]{navarro1997universal}
Navarro, J.~F., Frenk, C.~S., and White, S.~D. (1997).
\newblock A universal density profile from hierarchical clustering.
\newblock \emph{The Astrophysical Journal} 490, 493
\input{navarro1997universal}

\bibitem[{Nieuwenhuizen(2017)}]{nieuwenhuizen2017zwicky}
Nieuwenhuizen, T.~M. (2017).
\newblock How {Z}wicky already ruled out modified gravity theories without dark
  matter.
\newblock \emph{Fortschritte der Physik} 65, 1600050
\input{nieuwenhuizen2017zwicky}

\bibitem[{Nieuwenhuizen(2020)}]{nieuwenhuizen2020subjecting}
Nieuwenhuizen, T.~M. (2020).
\newblock Subjecting dark matter candidates to the cluster test.
\newblock \emph{Fluctuation and Noise {L}etters} 19, 2050016
\input{nieuwenhuizen2020subjecting}

\bibitem[{Nieuwenhuizen(2021)}]{nieuwenhuizen2021interior}
Nieuwenhuizen, T.~M. (2021).
\newblock The interior of hairy black holes in standard model physics.
\newblock \emph{arXiv preprint arXiv:2108.01422}
\input{nieuwenhuizen2021interior}

\bibitem[{Nieuwenhuizen(2023{\natexlab{a}})}]{nieuwenhuizen2023exact}
Nieuwenhuizen, T.~M. (2023{\natexlab{a}}).
\newblock Exact solutions for black holes with a smooth quantum core.
\newblock \emph{arXiv preprint arXiv:2302.14653}
\input{nieuwenhuizen2023exact}

\bibitem[{Nieuwenhuizen(2023{\natexlab{b}})}]{nieuwenhuizen2023solution}
Nieuwenhuizen, T.~M. (2023{\natexlab{b}}).
\newblock Solution of the dark matter riddle within standard model physics:
  {F}rom galaxies and clusters to cosmology.
\newblock \emph{arXiv preprint arXiv:2303.04637v1}
\input{nieuwenhuizen2023solution}

\bibitem[{Nieuwenhuizen(2024)}]{nieuwenhuizen2024aether}
Nieuwenhuizen, T.~M. (2024).
\newblock How the aether rescues the {L}orentz electron and imprints its
  {N}ewtonian and geodesic motion, and the equivalence principle.
\newblock \emph{submitted}
\input{nieuwenhuizen2024aether}

\bibitem[{Nieuwenhuizen et~al.(2021)Nieuwenhuizen, Limousin, and
  Morandi}]{nieuwenhuizen2021accurate}
Nieuwenhuizen, T.~M., Limousin, M., and Morandi, A. (2021).
\newblock Accurate modeling of the strong and weak lensing profiles for the
  galaxy clusters {A}bell 1689 and 1835.
\newblock \emph{The European Physical Journal Special Topics} 230, 1137--1148
\input{nieuwenhuizen2021accurate}

\bibitem[{Nieuwenhuizen and Morandi(2013)}]{nieuwenhuizen2013observations}
Nieuwenhuizen, T.~M. and Morandi, A. (2013).
\newblock Are observations of the galaxy cluster {A}1689 consistent with a
  neutrino dark matter scenario?
\newblock \emph{Monthly Notices of the Royal Astronomical Society} 434,
  2679--2683
\input{nieuwenhuizen2013observations}

\bibitem[{Oehm and Kroupa(2024)}]{oehm2024relevance}
Oehm, W. and Kroupa, P. (2024).
\newblock The relevance of dynamical friction for the {MW/LMC/SMC} triple
  system.
\newblock \emph{{U}niverse} 10, 143
\input{oehm2024relevance}

\bibitem[{Ou et~al.(2024)Ou, Eilers, Necib, and Frebel}]{ou2024dark}
Ou, X., Eilers, A.-C., Necib, L., and Frebel, A. (2024).
\newblock The dark matter profile of the {M}ilky {W}ay inferred from its
  circular velocity curve.
\newblock \emph{Monthly Notices of the Royal Astronomical Society} , stae034
\input{ou2024dark}

\bibitem[{Palunas and Williams(2000)}]{palunas2000maximum}
Palunas, P. and Williams, T. (2000).
\newblock Maximum disk mass models for spiral galaxies.
\newblock \emph{The Astronomical Journal} 120, 2884
\input{palunas2000maximum}

\bibitem[{Pandya et~al.(2024)Pandya, Zhang, Iyer, McGrath, Barro, Finkelstein
  et~al.}]{pandya2024galaxies}
Pandya, V., Zhang, H., Iyer, K.~G., McGrath, E., Barro, G., Finkelstein, S.~L.,
  et~al. (2024).
\newblock Galaxies going bananas: Inferring the 3d geometry of high-redshift
  galaxies with {{JWST}}-{CEERS}.
\newblock \emph{The Astrophysical Journal} 963, 54
\input{pandya2024galaxies}

\bibitem[{Pawlowski et~al.(2012)Pawlowski, Pflamm-Altenburg, and
  Kroupa}]{pawlowski2012vpos}
Pawlowski, M., Pflamm-Altenburg, J., and Kroupa, P. (2012).
\newblock The {VPOS}: a vast polar structure of satellite galaxies, globular
  clusters and streams around the {M}ilky {W}ay.
\newblock \emph{Monthly Notices of the Royal Astronomical Society} 423,
  1109--1126
\input{pawlowski2012vpos}

\bibitem[{Pawlowski et~al.(2014)Pawlowski, Famaey, Jerjen, Merritt, Kroupa,
  Dabringhausen et~al.}]{pawlowski2014co}
Pawlowski, M.~S., Famaey, B., Jerjen, H., Merritt, D., Kroupa, P.,
  Dabringhausen, J., et~al. (2014).
\newblock Co-orbiting satellite galaxy structures are still in conflict with
  the distribution of primordial dwarf galaxies.
\newblock \emph{Monthly Notices of the Royal Astronomical Society} 442,
  2362--2380
\input{pawlowski2014co}

\bibitem[{Peebles(1982)}]{peebles1982large}
[Dataset] Peebles, P. (1982).
\newblock Large scale background temperature and mass fluctuations due to scale
  invariant primeval perturbations
\input{peebles1982large}

\bibitem[{Penrose(1989)}]{roger1989emperor}
Penrose, R. (1989).
\newblock \emph{The emperor's new mind: concerning computers, minds, and the
  laws of physics} (Oxford University Press)
\input{roger1989emperor}

\bibitem[{Perivolaropoulos and Skara(2022)}]{perivolaropoulos2022challenges}
Perivolaropoulos, L. and Skara, F. (2022).
\newblock Challenges for $\lambda$cdm: {A}n update.
\newblock \emph{New Astronomy {R}eviews} , 101659
\input{perivolaropoulos2022challenges}

\bibitem[{Perlmutter et~al.(1999)Perlmutter, Aldering, Goldhaber, Knop, Nugent,
  Castro et~al.}]{perlmutter1999measurements}
Perlmutter, S., Aldering, G., Goldhaber, G., Knop, R., Nugent, P., Castro,
  P.~G., et~al. (1999).
\newblock Measurements of $\omega$ and $\lambda$ from 42 high-redshift
  supernovae.
\newblock \emph{The Astrophysical Journal} 517, 565
\input{perlmutter1999measurements}

\bibitem[{Peskin and Schroeder(2018)}]{peskin2018introduction}
Peskin, M.~E. and Schroeder, D.~V. (2018).
\newblock \emph{An introduction to quantum field theory} (CRC Press)
\input{peskin2018introduction}

\bibitem[{Pignol et~al.(2015)Pignol, Clement, Guigue, Rebreyend, and
  Voirin}]{pignol2015constraints}
Pignol, G., Clement, B., Guigue, M., Rebreyend, D., and Voirin, B. (2015).
\newblock Constraints on dark photon dark matter using {V}oyager magnetometric
  survey.
\newblock \emph{arXiv preprint arXiv:1507.06875}
\input{pignol2015constraints}

\bibitem[{Pitrou et~al.(2018)Pitrou, Coc, Uzan, and
  Vangioni}]{pitrou2018precision}
Pitrou, C., Coc, A., Uzan, J.-P., and Vangioni, E. (2018).
\newblock Precision {B}ig {B}ang nucleosynthesis with improved helium-4
  predictions.
\newblock \emph{Physics Reports} 754, 1--66
\input{pitrou2018precision}

\bibitem[{Reucroft(2014)}]{reucroft2014galactic}
Reucroft, S. (2014).
\newblock Galactic charge.
\newblock \emph{arXiv preprint arXiv:1409.3096}
\input{reucroft2014galactic}

\bibitem[{Rich(2009)}]{rich2009fundamentals}
Rich, J. (2009).
\newblock \emph{Fundamentals of cosmology} (Springer Science \& Business Media)
\input{rich2009fundamentals}

\bibitem[{Riess et~al.(1998)Riess, Filippenko, Challis, Clocchiatti, Diercks,
  Garnavich et~al.}]{riess1998observational}
Riess, A.~G., Filippenko, A.~V., Challis, P., Clocchiatti, A., Diercks, A.,
  Garnavich, P.~M., et~al. (1998).
\newblock Observational evidence from supernovae for an accelerating {U}niverse
  and a cosmological constant.
\newblock \emph{The Astronomical Journal} 116, 1009
\input{riess1998observational}

\bibitem[{Rodrigues et~al.(2018)Rodrigues, Marra, del Popolo, and
  Davari}]{rodrigues2018absence}
Rodrigues, D.~C., Marra, V., del Popolo, A., and Davari, Z. (2018).
\newblock Absence of a fundamental acceleration scale in galaxies.
\newblock \emph{Nature Astronomy} 2, 668--672
\input{rodrigues2018absence}

\bibitem[{Roshan et~al.(2021)Roshan, Ghafourian, Kashfi, Banik, Haslbauer,
  Cuomo et~al.}]{roshan2021fast}
Roshan, M., Ghafourian, N., Kashfi, T., Banik, I., Haslbauer, M., Cuomo, V.,
  et~al. (2021).
\newblock Fast galaxy bars continue to challenge standard cosmology.
\newblock \emph{Monthly Notices of the Royal Astronomical Society} 508,
  926--939
\input{roshan2021fast}

\bibitem[{Rubin and Ford~Jr(1970)}]{rubin1970rotation}
Rubin, V.~C. and Ford~Jr, W.~K. (1970).
\newblock Rotation of the {A}ndromeda nebula from a spectroscopic survey of
  emission regions.
\newblock \emph{The Astrophysical Journal} 159, 379
\input{rubin1970rotation}

\bibitem[{Rycroft et~al.(2000)Rycroft, Israelsson, and
  Price}]{rycroft2000global}
Rycroft, M., Israelsson, S., and Price, C. (2000).
\newblock The global atmospheric electric circuit, solar activity and climate
  change.
\newblock \emph{Journal of Atmospheric and Solar-Terrestrial Physics} 62,
  1563--1576
\input{rycroft2000global}

\bibitem[{Salucci and Burkert(2000)}]{salucci2000dark}
Salucci, P. and Burkert, A. (2000).
\newblock Dark matter scaling relations.
\newblock \emph{The Astrophysical Journal} 537, L9
\input{salucci2000dark}

\bibitem[{Sancisi(2004)}]{sancisi2004visible}
Sancisi, R. (2004).
\newblock The visible matter-dark matter coupling.
\newblock In \emph{Symposium International Astronomical Union} (Cambridge
  University Press), vol. 220, 233--240
\input{sancisi2004visible}

\bibitem[{Sawala et~al.(2023)Sawala, Cautun, Frenk, Helly, Jasche, Jenkins
  et~al.}]{sawala2023milky}
Sawala, T., Cautun, M., Frenk, C., Helly, J., Jasche, J., Jenkins, A., et~al.
  (2023).
\newblock The {M}ilky {W}ay’s plane of satellites is consistent with
  {$\Lambda$CDM}.
\newblock \emph{Nature Astronomy} 7, 481--491
\input{sawala2023milky}

\bibitem[{Schlaufman et~al.(2018)Schlaufman, Thompson, and
  Casey}]{schlaufman2018ultra}
Schlaufman, K.~C., Thompson, I.~B., and Casey, A.~R. (2018).
\newblock An ultra metal-poor star near the hydrogen-burning limit.
\newblock \emph{The Astrophysical Journal} 867, 98
\input{schlaufman2018ultra}

\bibitem[{Secrest et~al.(2022)Secrest, von Hausegger, Rameez, Mohayaee, and
  Sarkar}]{secrest2022challenge}
Secrest, N.~J., von Hausegger, S., Rameez, M., Mohayaee, R., and Sarkar, S.
  (2022).
\newblock A challenge to the standard cosmological model.
\newblock \emph{The Astrophysical journal letters} 937, L31
\input{secrest2022challenge}

\bibitem[{Sharma et~al.(2022)Sharma, Salucci, and van~de
  Ven}]{sharma2022observational}
Sharma, G., Salucci, P., and van~de Ven, G. (2022).
\newblock Observational evidence of evolving dark matter profiles at $z\le 1$.
\newblock \emph{Astronomy \& Astrophysics} 659, A40
\input{sharma2022observational}

\bibitem[{Shelest and Lelli(2020)}]{shelest2020spirals}
Shelest, A. and Lelli, F. (2020).
\newblock From spirals to lenticulars: Evidence from the rotation curves and
  mass models of three early-type galaxies.
\newblock \emph{Astronomy \& Astrophysics} 641, A31
\input{shelest2020spirals}

\bibitem[{Sikivie(1983)}]{sikivie1983experimental}
Sikivie, P. (1983).
\newblock Experimental tests of the ``invisible" axion.
\newblock \emph{Physical {R}eview {L}etters} 51, 1415
\input{sikivie1983experimental}

\bibitem[{Spite and Spite(1982)}]{spite1982abundance}
Spite, F. and Spite, M. (1982).
\newblock Abundance of lithium in unevolved halo stars and old disk
  stars-interpretation and consequences.
\newblock \emph{Astronomy and Astrophysics} 115, 357--366
\input{spite1982abundance}

\bibitem[{Tisserand et~al.(2007)Tisserand, Le~Guillou, Afonso, Albert,
  Andersen, Ansari et~al.}]{tisserand2007limits}
Tisserand, P., Le~Guillou, L., Afonso, C., Albert, J., Andersen, J., Ansari,
  R., et~al. (2007).
\newblock Limits on the {M}acho content of the galactic halo from the {EROS}-2
  {S}urvey of the {M}agellanic clouds.
\newblock \emph{Astronomy \& Astrophysics} 469, 387--404
\input{tisserand2007limits}

\bibitem[{Vagnozzi(2023)}]{vagnozzi2023seven}
Vagnozzi, S. (2023).
\newblock Seven hints that early-time new physics alone is not sufficient to
  solve the hubble tension.
\newblock \emph{{U}niverse} 9, 393
\input{vagnozzi2023seven}

\bibitem[{van~der Wel et~al.(2014)van~der Wel, Chang, Bell, Holden, Ferguson,
  Giavalisco et~al.}]{van2014geometry}
van~der Wel, A., Chang, Y.-Y., Bell, E., Holden, B., Ferguson, H., Giavalisco,
  M., et~al. (2014).
\newblock Geometry of star-forming galaxies from sdss, 3d-hst, and candels.
\newblock \emph{The Astrophysical Journal Letters} 792, L6
\input{van2014geometry}

\bibitem[{Veltman(1975)}]{veltman1975cosmology}
Veltman, M. (1975).
\newblock Cosmology and the higgs mass.
\newblock \emph{Physical {R}eview {L}etters} 34, 777
\input{veltman1975cosmology}

\bibitem[{Verlinde(2011)}]{verlinde2011origin}
Verlinde, E. (2011).
\newblock On the origin of gravity and the laws of {N}ewton.
\newblock \emph{Journal of High Energy Physics} 2011, 1--27
\input{verlinde2011origin}

\bibitem[{Verlinde(2017)}]{verlinde2017emergent}
Verlinde, E. (2017).
\newblock Emergent gravity and the dark {U}niverse.
\newblock \emph{SciPost Physics} 2, 016
\input{verlinde2017emergent}

\bibitem[{Vogelsberger et~al.(2014)Vogelsberger, Genel, Springel, Torrey,
  Sijacki, Xu et~al.}]{vogelsberger2014introducing}
Vogelsberger, M., Genel, S., Springel, V., Torrey, P., Sijacki, D., Xu, D.,
  et~al. (2014).
\newblock Introducing the {I}llustris {P}roject: simulating the coevolution of
  dark and visible matter in the {U}niverse.
\newblock \emph{Monthly Notices of the Royal Astronomical Society} 444,
  1518--1547
\input{vogelsberger2014introducing}

\bibitem[{Weinberg(1972)}]{weinberg1972gravitation}
Weinberg, S. (1972).
\newblock \emph{{G}ravitation and cosmology: principles and applications of the
  general theory of relativity} (John Wiley \& Sons)
\input{weinberg1972gravitation}

\bibitem[{Weinberg(1978)}]{weinberg1978new}
Weinberg, S. (1978).
\newblock A new light boson?
\newblock \emph{Physical {R}eview {L}etters} 40, 223
\input{weinberg1978new}

\bibitem[{Wilczek(1978)}]{wilczek1978problem}
Wilczek, F. (1978).
\newblock Problem of strong {P} and {T} invariance in the presence of
  instantons.
\newblock \emph{Physical {R}eview {L}etters} 40, 279
\input{wilczek1978problem}

\bibitem[{Zel'dovich(1967)}]{zel1967cosmological}
Zel'dovich, Y.~B. (1967).
\newblock Cosmological constant and elementary particles.
\newblock \emph{ZhETF Pisma Redaktsiiu} 6, 883
\input{zel1967cosmological}

\bibitem[{Zel'{d}ovich(1968)}]{zel1968cosmological}
Zel'{d}ovich, Y.~B. (1968).
\newblock The cosmological constant and the theory of elementary particles.
\newblock \emph{Soviet Physics Uspekhi} 11, 381
\input{zel1968cosmological}

\bibitem[{Zhang et~al.(2019)Zhang, Primack, Faber, Koo, Dekel, Chen
  et~al.}]{zhang2019evolution}
Zhang, H., Primack, J.~R., Faber, S., Koo, D.~C., Dekel, A., Chen, Z., et~al.
  (2019).
\newblock The evolution of galaxy shapes in candels: from prolate to discy.
\newblock \emph{Monthly Notices of the Royal Astronomical Society} 484,
  5170--5191
\input{zhang2019evolution}

\bibitem[{Zwicky(1933)}]{zwicky1933rotverschiebung}
Zwicky, F. (1933).
\newblock Die {R}otverschiebung von extragalaktischen {N}ebeln.
\newblock \emph{Helvetica {P}hysica {A}cta} 6, 110--127
\input{zwicky1933rotverschiebung}

\end{thebibliography}
\end{document}